\documentclass[aps,twocolumn,amssymb]{revtex4}
\usepackage{helvet,times}
\usepackage{bm,textcomp}
\usepackage{float}
\usepackage{amssymb,amsmath}
\usepackage{graphicx}

\def\beq{\begin{equation}}
\def\eeq{\end{equation}}
\def\imagetop#1{\vtop{\null\hbox{#1}}}
\begin{document}
\title{On the modeling of endocytosis in yeast}

\author{T. Zhang$^1$, R. Sknepnek$^{1,2,3}$, M. J. Bowick$^1$, and J. M. Schwarz$^1$}
\email{jschwarz@physics.syr.edu}
\affiliation{$^1$ Department of Physics, Syracuse University, Syracuse, NY 13244, USA}
\affiliation{$^2$ Division of Physics, University of Dundee, Dundee, DD1 4HN, UK}
\affiliation{ $^3$ Division of Computational Biology, University of Dundee, Dundee, DD1 4HN, UK }

\begin{abstract}
{The cell membrane deforms during endocytosis to surround extracellular material and draw it into the cell.  Experiments on endocytosis in yeast all agree that (i) actin polymerizes into a network of filaments exerting active forces on the membrane to deform it and (ii) the large scale membrane deformation is tubular in shape. There are three competing proposals, in contrast, for precisely how the actin filament network organizes itself to drive the deformation. We use variational approaches and numerical simulations to address this competition by analyzing a meso-scale model of actin-mediated endocytosis in yeast.  The meso-scale model breaks up the invagination process into three stages: (i) initiation, where clathrin interacts with the membrane via adaptor proteins, (ii) elongation, where the membrane is then further deformed by polymerizing actin filaments, followed by (iii) pinch-off.  Our results suggest that the pinch-off mechanism may be assisted by a pearling-like instability. We rule out two of the three competing proposals for the organization of the actin filament network during the elongation stage. These two proposals could possibly be important in the pinch-off stage, however, where additional actin polymerization helps break off the vesicle.   Implications and comparisons with earlier modeling of endocytosis in yeast are discussed.}

\end{abstract}

\maketitle

\section{Introduction}

Endocytosis is the process by which extracellular agents are ingested by the cell as a result of the cell membrane surrounding and engulfing them~\cite{lodish}. The membrane then pinches off to form a vesicle that encloses the now intracellular material. Fig. 1 presents an electron micrograph image of a deformed cell membrane near pinch-off in S. cerevisiae~\cite{kukulski}.  Experiments have identified a handful of core proteins, though there are upwards of 50 proteins participating in the endocytotic machinery~\cite{mcmahon,perrais,lemmon}. Live cell imaging of these fluorescently labeled core proteins provide us with a sequence of events for the endocytotic machinery~\cite{kaksonen,taylor}. Though the composition and time-line of the endocytotic machinery is known, in yeast, there are competing proposals about how these few core proteins interact with the cell membrane to deform it into a vesicle~\cite{endo1b,endo3,endo4}. We address these competing {\it qualitative} proposals by {\it quantitatively} comparing them. 
\vspace{-0.25cm}
\begin{figure}[h]
\begin{center}
\includegraphics[scale=0.75]{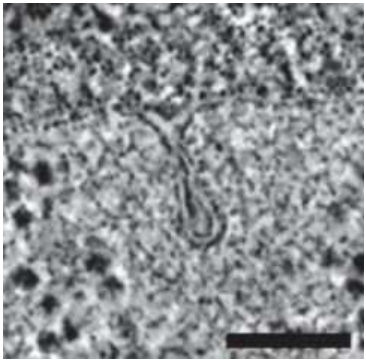}
\caption{An electron micrograph image of a deformed membrane during endocytosis in S. cerevisiae. The image is reprinted with permission from Ref. 2. The scale bar is $100\,\,nm$. }
\end{center}
\end{figure}

\begin{figure*}[t]
\centering
\begin{tabular}[t]{cccc}
\imagetop{\includegraphics[scale=0.22]{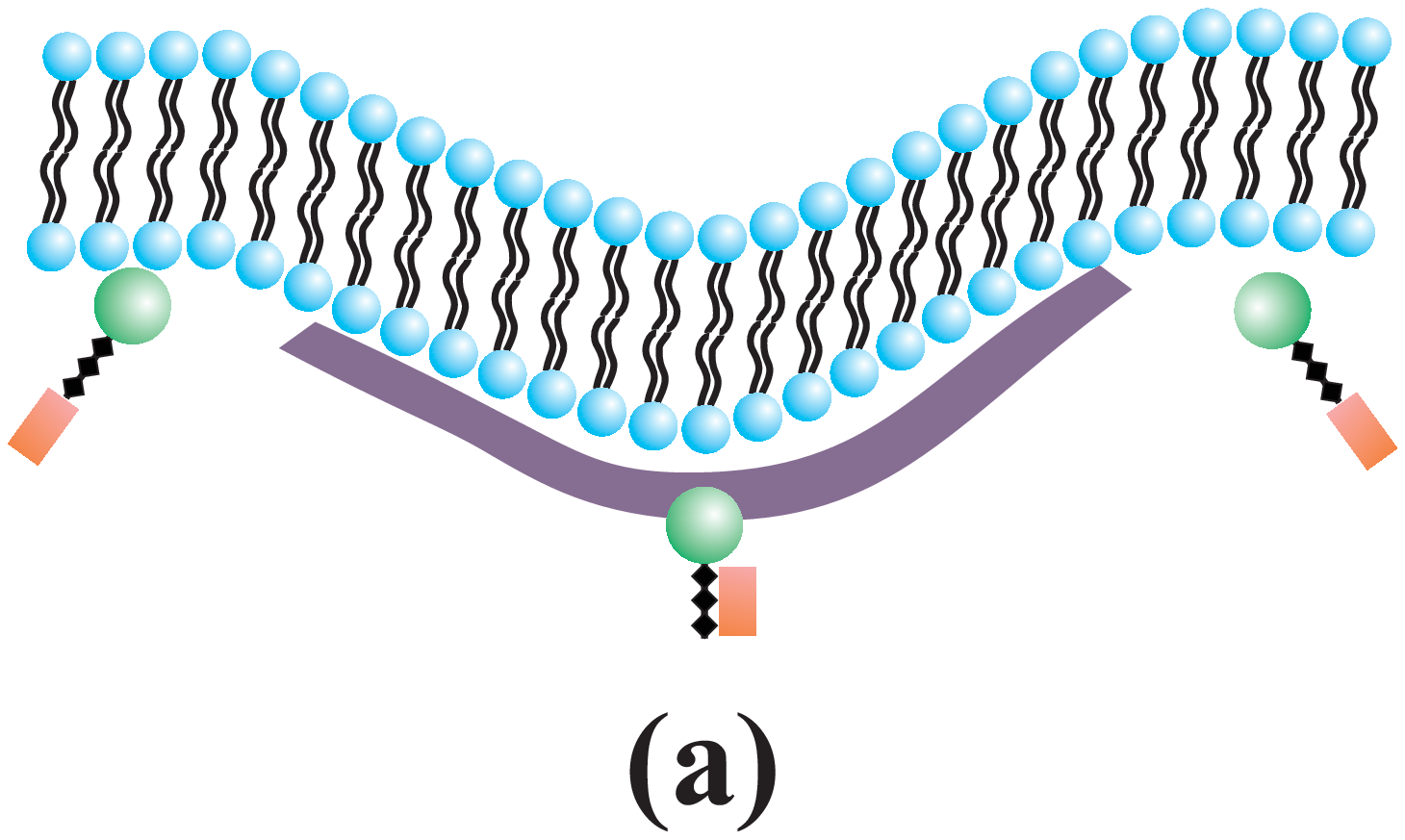}} &
\imagetop{\includegraphics[scale=0.22]{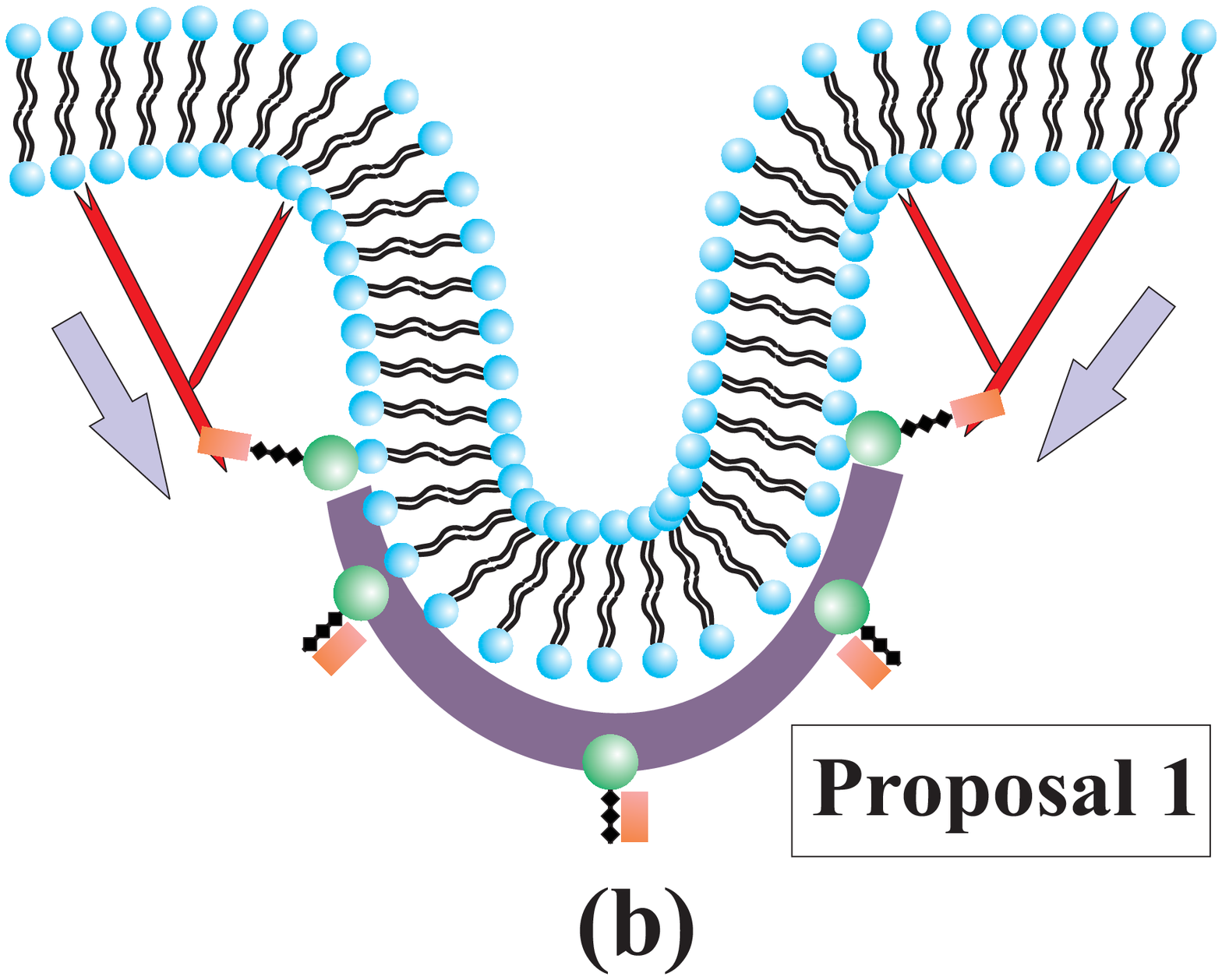}} &
\imagetop{\includegraphics[scale=0.22]{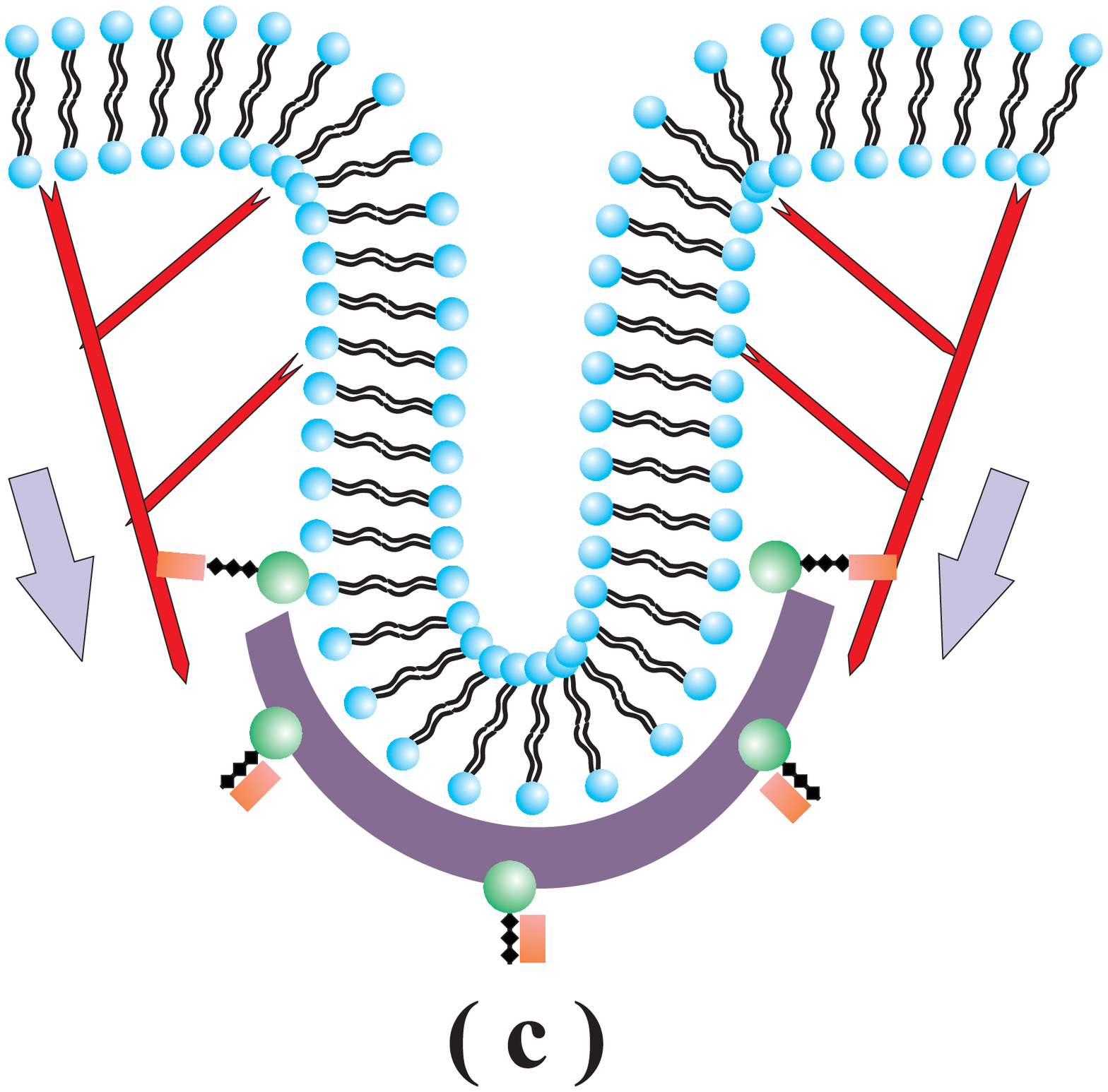}} &
\imagetop{\includegraphics[scale=0.22]{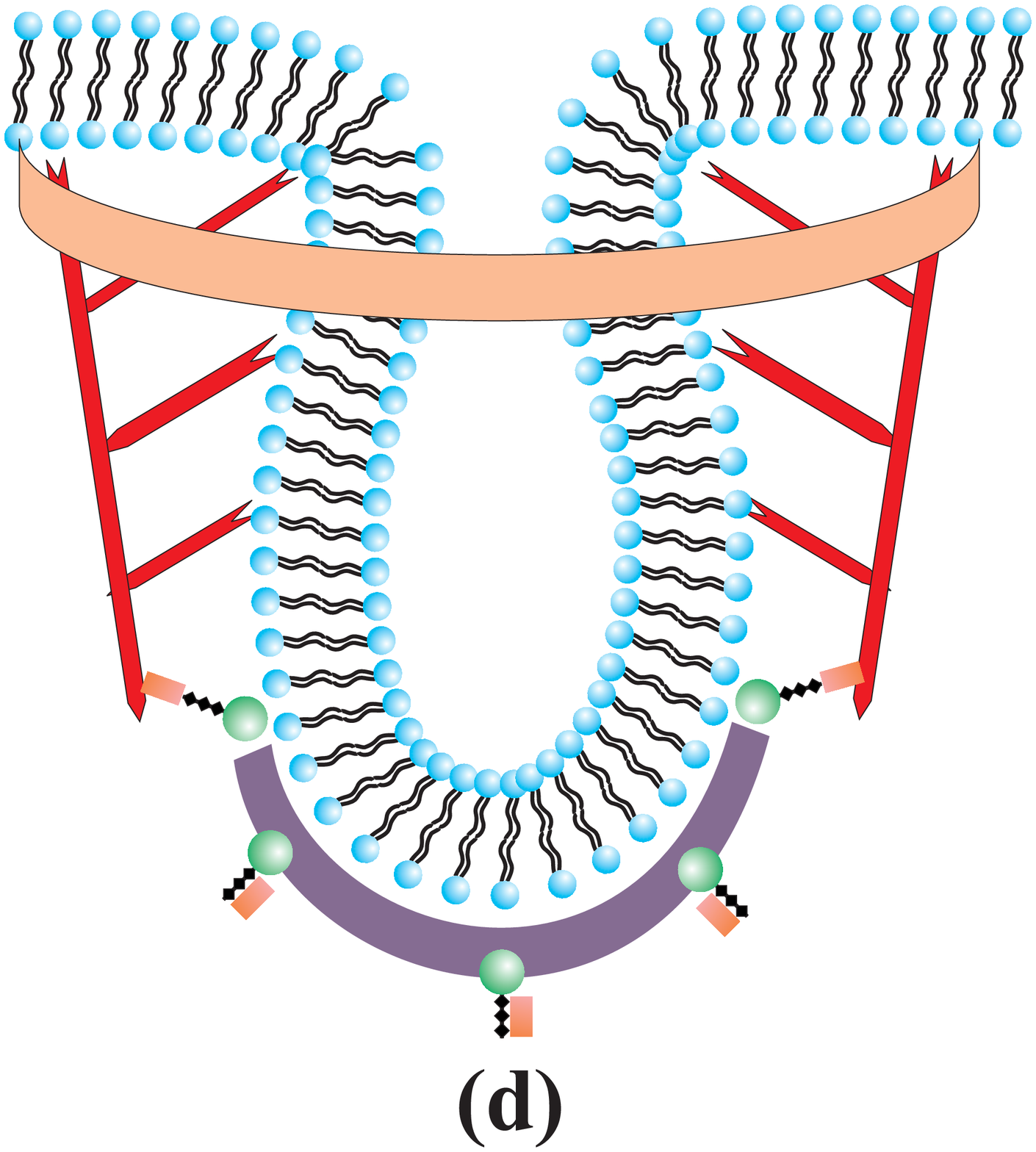}} 
\end{tabular}
\caption{Schematic for endocytosis in yeast using Proposal 1 for the actin filament organization: (a) Clathrin (purple) attaches to the membrane (black/blue) via proteins Sla1 and Ent1/2 (not depicted here) and the protein Sla2 (green/brown) is recruited near the clathrin. (b) Actin (red) attaches to the membrane near the edge of the clathrin ``bowl'' via Sla2 and lengthens due to polymerization to initiate tube formation. (c) Actin continues to polymerize and lengthen the tube. (d) BAR proteins (orange) become prominent and surround part of the tube (and the actin). The grey arrows denote the direction of the actin force on the membrane. Note that potential additional actin filaments rooted in the surrounding cytoskeleton and extending towards the invagination site not been drawn.}
\end{figure*}

According to experiments, the sequence of events in the endocytotic machinery in yeast is as follows~\cite{review,endo1}. Clathrin is recruited to the invagination site~\cite{clathrin1}, along with adaptor proteins, such as Sla1 and Ent1/2~\cite{lemmon}. Sla1 and Ent1/2 proteins bind the clathrin to the membrane, while Sla2 proteins bind actin filaments to the membrane~\cite{lemmon}.  Another protein, WASp, is also recruited to the site. WASp is an activator for the branching agent Arp2/3, enabling a branched actin filament network to be generated near the invagination site~\cite{kaksonen}. The growth of this network drives membrane tube formation. BAR proteins eventually become prominent and help facilitate pinch-off of the membrane~\cite{BAR}. Fig. 2 illustrates this process using what will turn out to be Proposal 1 for the organization of the actin.  The initial invagination due to clathrin and other adaptor proteins takes about one to two minutes. The time for the tube to form and pinch-off, in contrast, takes only about 10-15 seconds.  The length-to-radius ratio of the tube before pinch-off is typically 7-10~\cite{kukulski} (Fig. 1).

\begin{figure}[t]
\centering
\includegraphics[scale=0.25]{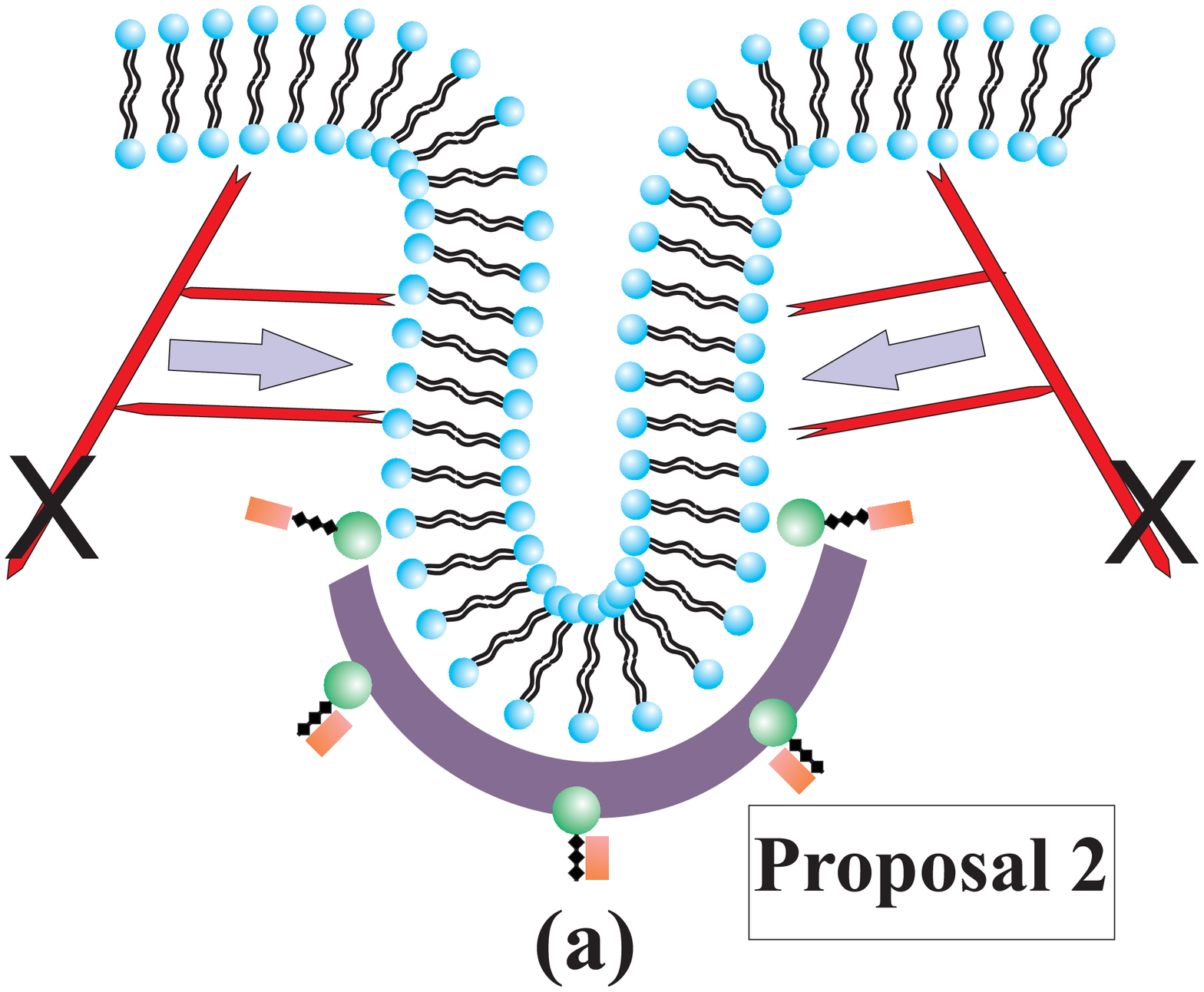}
\includegraphics[scale=0.25]{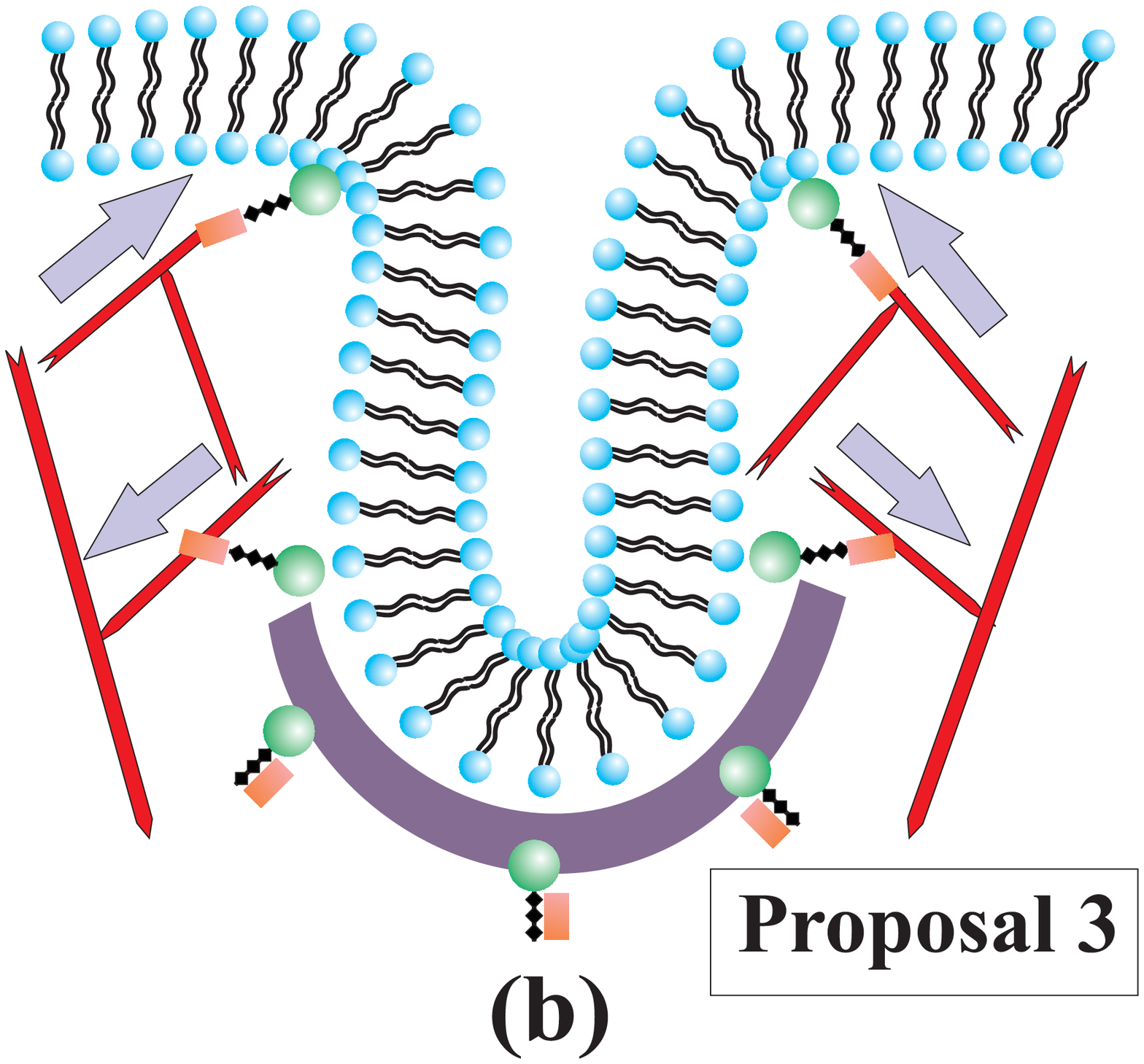}
\caption{(a) Schematic depicts Proposal 2, where the actin filaments are tethered to the rest of the cytoskeleton, as denoted by the two black Xs, and polymerize inward towards the invagination site. (b) Schematic represents Proposal 3, where there are two local anchoring regions such that two actin networks form to drive tube formation. The grey arrows, again, denote the direction of the actin force on the membrane. }
\end{figure}

The role of clathrin may be somewhat clear. The spontaneous curvature of individual clathrin molecules presumably helps initiate membrane deformation as they indirectly attach to the membrane via adaptor proteins with the initial deformation being rather small, in contrast to mammalian cells. How the actin filament network reshapes the membrane, on the other hand, is even less clear since there are currently three competing proposals put forth by the biologists as to how this is done. The first proposal (Proposal 1) argues that the barbed/plus ends of polymerizing actin filaments are oriented towards the flat part of the membrane with the pointed/minus ends anchored just above the clathrin bowl~\cite{endo1b}. A second proposal (Proposal 2) argues that a collar-like structure of plus end filaments anchored to the rest of the cytoskeleton and oriented towards the neck of the deformation to elongate it and drive the pinch-off~\cite{endo3}. A third proposal (Proposal 3) suggests that there are two regions of attachment of the actin filaments to the membrane such that two branched actin networks are generated~\cite{endo4}. The two networks repel each other as they grow because they cannot interpenetrate and, therefore, drive tube formation. See Fig. 2b for a schematic of Proposal 1 and Fig. 3 for a schematic for Proposals 2 and 3.

Each of these three proposals for endocytosis in yeast assumes its own respective organization for the actin filament network. Can any of these models be ruled out on the basis that they do not provide the forces required to deform the cell membrane that is consistent with observations? And what about the mechanism for pinch-off?  Is there any underlying instability, or is there a more engineered approach with the pinch-off occuring at some fixed distance from the top of the invagination? We approach clathrin-initiated endoctyosis in yeast by breaking up the sequence of fluid membrane deformations into three stages: (i) initiation, (ii) elongation, and (iii) pinch-off.  In the process we identify a possible mechanism that could assist in the pinch-off via a pearling-like instability, where surface tension competes with bending energy in cylindrical vesicles such that, for long enough cylinders, it is energetically favorable for the cylinder to break up into spheres. An instability driven mechanism is potentially powerful given the ubiquitousness of endocytosis. 

This proposed pearling-like instability mechanism will be contrasted with a competing pinch-off mechanism put forth in Refs.~\cite{endo1b,endo1c}. In this model, the cell membrane is also modeled as a fluid membrane---a two-component fluid membrane, where one component consists of the nonscission region and the other component consisting of the scission region such. Physically, the two regions correspond to hydrolyzed PIP$_2$ and non-hydrolyzed PIP$_2$ regions. The increasing interfacial line tension between the two components drives the pinch-off with the pinch-off distance always occuring at some fixed distance from the top of the invagination site by construction, if you will. We will ultimately compare and contrast our model with this earlier model and compare our model with another more recent model in which the cell membrane is modeled as an elastic membrane with a non-zero shear modulus~\cite{bayly}. 

In mammalian cells, the two key players in endocytosis are clathrin and dynamin, a motor protein that drives the pinch-off, which is very different from yeast. One difference between yeast and mammalian cells is the presence of a cell wall in yeast cells. This wall is needed to prevent lysis due to an internal turgor pressure, which can be as large $10^6$ Pa.  It has been speculated that the presence of the turgor pressure biases the use of F-actin as an invagination tool~\cite{ayscough}.  We will discuss the implications of turgor pressure for our model throughout this work.

The paper is organized as follows. Section II introduces our cell membrane modelling approach to endocytosis, which is divided up into three stages as mentioned previously, while Section III presents the resulting cell membrane configurations for each stage.  We conclude with Section IV by framing our results and discussing their implications.

\section{Model and methods} 
We model the cell membrane as an ideal two-dimensional
surface residing in three-dimensional Euclidean space. This surface
represents the neutral surface of the physical membrane and approximately
corresponds to the contact between the two leaflets of the lipid bilayer~\cite{safran}.
Mathematically, this surface is parametrized by curvilinear coordinates
$\left(\alpha_{1},\alpha_{2}\right)$ and is described by radius vectors
$\vec{r}=\vec{r}\left(\alpha_{1},\alpha_{2}\right)\equiv\left(x\left(\alpha_{1},\alpha_{2}\right),y\left(\alpha_{1},\alpha_{2}\right),z\left(\alpha_{1},\alpha_{2}\right)\right)$. Using standard methods of differential geometry of surfaces~\cite{carmo}, we then define the mean ($H$) and Gaussian $\left(K\right)$
curvatures of the surface. 

The energy of a bare membrane depends on its curvature~\cite{helfrich} and can be written as
\begin{equation}
E_{bare}=\int dS\left[2\kappa\left(H-C_{0}\right)^{2}+\kappa_G K+\sigma\right] +p\int dV,\label{eq:Helfriech}
\end{equation}
where $C_{0}$ is a spontaneous curvature, $\kappa$ is the membrane
bending rigidity, $\kappa_G$ is the saddle-splay modulus, $\sigma$ represents the surface tension, and $dS$
represents the area of an infinitesimal element of the surface. Finally, $p$ represents the turgor pressure present in yeast cells with $dV$ the infinitesimal volume element for any deformation from a flat surface.

Beginning with the above energy functional, which models the energy of the deformations of a bare cellular membrane, we systematically incorporate new physics associated with the three stages of endocytosis by allowing the parameters to be component-dependent or by adding new terms to the energy. 

\subsection{Initiation stage}
Clathrin is one of the first proteins recruited to the endocytic site. Each clathrin molecule is a nonplanar triskelion that can pucker in the center~\cite{kirchhausen}. Clathrin molecules bind together to form a basket-like structure as a result of the intrinsic curvature of the molecules. Clathrin molecules, however, require adaptor proteins, such as Sla1 and Ent1/2, to bind to the membrane~\cite{wendland,lemmon}.  The binding process induces curvature in the membrane.  The membrane rigidity may also be affected by protein binding. In fact, membrane rigidities depend on several factors, such as membrane lipid and protein composition, to account for the range of values (tens of $k_BT$) that is reported in the literature~\cite{boal}.

We encode the effect of the clathrin binding, via Sla1 and Ent1/2, to the cell membrane with effective parameters characterizing the model of the cell membrane. The clathrin indirectly binding to one side of the membrane induces spontaneous curvature in the membrane. Since the clathrin indirectly binds only to part of the membrane, we study a two-component membrane, one component denoting the bare membrane and the other denoting the part of the membrane to which the clathrin is indirectly attached with non-zero spontaneous curvature. We also vary the bending rigidity of the part of the membrane to which the clathrin is indirectly attached. 

The energy functional of the membrane for this initiation stage is given by~\cite{helfrich}
\beq
E_{init}=
\sum_{i=1,2}\int dS_i\left[2\kappa_i \left(H_i-C_{0i} \right)^2+\kappa_{G_i} K_i+\sigma_i\right] +p\int dV_i
\eeq
where $i=1$ denotes the Sla1/Ent1/2-bound membrane and $i=2$ denotes the bare membrane.  

\subsection{Elongation stage}

We now ask how the emergent actin network exerts additional deformations/forces on the membrane following the initiation of endocytosis. As shown in Fig. 2, the protein Sla2 is recruited to the invagination site before actin assembly. Sla2 binds to the clathrin and the membrane near the clathrin, but, according to Ref.~\cite{lemmon}, Sla2 bound to the membrane near the top of the clathrin basket binds actin filaments. These Sla2 binding sites provide localized binding/anchoring of the actin filaments to the cell membrane since these Sla2 molecules are near the top of the clathrin basket, i.e. "butting" up against the elastic clathrin basket.  Assuming the minus end of the actin filament anchors to the cell membrane via Sla2, the plus end then polymerizes upward and ratchets against the cell membrane. The asymmetry between anchoring at the minus end and ratcheting at the plus end provides a time-averaged force to invaginate the membrane further into the cell. Actin filament nucleation via Arp2/3 increases this force.  

We, therefore, model the actin filament network as an applied force on the membrane localized at these Sla2 anchoring points.  The magnitude of this force is related to the total number of actin filaments participating in the network, and this number has been computed based on a combination of experimental data and kinetic modelling~\cite{berro1,berro2}. We use the final configuration of the emergent actin network to determine the force applied to the membrane.  Given the observed tubular structure of the deformation, this actin force is assumed to be axisymmetric
with constant components in the radial and $-z$ (downward) directions, i.e. $\vec{F}_{act}=F_{\rho}\vec{e}_{\rho}+F_{z}\vec{e}_{z}$. The actin force is imposed by adding
a linear potential of the form $V_{act}\left(\rho,\varphi,z\right)=-\left(F_{\rho}\rho+F_{z}z\right)$
to the energy for the part of the membrane to which the force is locally applied. The energy functional for the elongation stage is 
\begin{equation}
\begin{split}
E_{elong}=E_{init}+\int d\vec{r}\,[V_{act}\left(\rho,\varphi,z\right)g\left(\rho,\varphi,z\right)\\
+V_{ster}\left(\rho,\varphi,z\right)+V_{pin}\left(\rho,\varphi,z\right)],
\end{split}
\end{equation}
where $g\left(\rho,\varphi,z\right)=1$ for the region over which the actin force is applied and zero otherwise. To distinguish between Proposals 1, 2, and 3, we explore different anchoring regions and different ratios of the force components. Note also that $V_{ster}\left(\rho,\varphi,z<0\right)=\infty$ for $\rho>R_{ap}$ and zero otherwise.  This models the accumulation of the yeast actin cytoskeleton just beneath the cell membrane and near the tubular invagination as it emerges~\cite{endo3,moseley}.  The $\rho>R_{ap},z>0$ region acts as
a ``reservoir'' for tube growth. We, therefore, impose an additional quadratic potential, $V_{pin}(\rho,z)=\frac{1}{2}\beta z^2$ for $\rho>R_{ap}$, where $\beta$ is chosen so that the membrane outside this region remains flat.

{\it Pinchoff stage:} Experiments indicate that the BAR proteins dominate in this last stage, {\it after} the tubular-like deformation forms~\cite{endo1c,kukulski}. This observation is rather perplexing since BAR proteins, which themselves are curved, can, sense and generate spontaneous curvature in the cell membrane~\cite{sorre}. In other words, why does actin play the dominant role in generating the tubular-like deformation and not the BAR proteins? We suggest that once a tubular-like deformation occurs, the BAR proteins surround and confine the tube-plus-actin filament network near the top of the invagination site (see Fig. 2(d)) to stop actin polymerization.  The BAR proteins only sense curvature here, not generate it. Since actin polymerization is driven by a ratcheting effect in a spatially fluctuating membrane, when these spatially fluctuations are supressed, actin polymerization stops. When the polymerization stops, no more membrane material can become part of the tube. In other words, the membrane tube area remains constant. With the BAR proteins confining the top part of the actin filament network against the membrane to couple the network to the membrane, we introduce an additional energetic term to the system. Now that the actin network has developed, we model it as an underlying elastic network of springs.  Because the actin network is now connected to the membrane (as opposed to ratcheting against it in the elongation stage), the filament tips of the spring network depends on the configuration of the membrane.  As with any elastic network coupled to a fluid membrane, the BAR protein-plus-actin filament contribution to the energy of a now cylindrical membrane is~\cite{brown}  
\begin{equation}
E_{BAR+actin}=\sum_{<i,j>} \frac{\mu}{2}[\vec{r}_i-\vec{r}_j]^2,
\end{equation}
where $\mu$ is the spring constant and $i,j$ denotes the meshwork coordinates of the springs on the surface of the membrane. Since the tube has now been formed, we will study a cylinder membrane described by Eq. (1) with this additional energy, $E_{BAR+actin}$.  This energy will turn out to raise the surface tension of the membrane. This calculation will also suggest a new pearling instability pinch-off mechanism for endocytosis in yeast.\\

{\it Methods:} We utilize both analytical and numerical techniques to study the above model. On the numerical side, we use simulated annealing Monte Carlo (MC) simulations to identify low-energy structures. 
In our simulations we represent the membrane using standard techniques
for constructing discrete surface triangulations \cite{gompper2004}. The discrete version of the bending energy in Eq. (2) is then implemented using expressions introduced by Brakke\cite{Brakke1992}. The mean curvature at vertex $i$ is given as a $H_i = \frac{1}{2}\frac{\mathbf{F}_i\cdot\mathbf{N}_i}{\mathbf{N}_i\cdot\mathbf{N}_i}$ with $\mathbf{F}_i$ being the gradient of area and $\mathbf{N}_i$ being gradient of volume calculated with respect to the coordinates of vertex $i$. The bending energy is then given as $E_{bend}=(H-C_{0})^2 A_i/3$, where $A_i$ is the total area of triangles sharing vertex $i$. The spontaneous curvature $C_{0}$ is chosen according to the region of the membrane to which the vertex belongs. Surface tension is computed as an energy penalty to change the reference area of the membrane. Reference area $A_0$ is chosen to be that of the initial flat configuration. The energy associated with changes of the surface area is then $E_{surface} = \sigma\left|A - A_0\right|$, where $A$ is the area associated with a vertex and computed as one third of the sum of areas of all triangles sharing that vertex.

To ensure that we simulate a fluid membrane,
each MC step involves two steps \cite{gompper2004}: \emph{i}) displace
a vertex in a direction chosen at random uniformly from a cube $\left[-0.05l_{0},0.05l_{0}\right]^{3}$, where $l_{0}$ is the average edge length of the initial triangulation, and
\emph{ii}) flip an edge on a rhombus. This flip removes an edge shared by two triangles
and reconnects it so that it spans the opposite, previously unattached,
vertices \cite{kazakov,billoire}. Moves are accepted or rejected according to the Metropolis
algorithm. The sweeps are continued until the total energy does not
change with some prescribed precision. Different random number generator
seeds are used to ensure the reproducibility of the lowest energy
configurations. A typical run with $N_{v}\approx3.5\times10^{3}$
vertices consists of $\sim10^{6}$ MC sweeps, with a sweep consisting
of attempted moves of each vertex and attempted flips of each edge.
Any moves or edge flips leading to unphysical self-intersection
of the triangulation are rejected. This is achieved by endowing
each vertex with a hard core of diameter $b=0.9l_{0}$, and each
edge with a tethering potential \cite{gompper2004} with maximum length
$l_{max}=1.4l_{0}$ such that $l_{max}/b\approx1.55$. These
values are chosen in accordance with Refs. \cite{gompper2000,kohyama}
to be tight enough to prevent edge crossings but still allow edge-flips,
thus ensuring fluidity. Finally, the actin forces are applied to the vertices. 
\section{Results}
\subsection{Initiation stage with clathrin}

To analyze the equilibrium shapes of the membrane in the initiation stage, we use the Monge representation such
 that each coordinate on the membrane in three-dimensional space is parameterized by two planar coordinates $x$ and $y$ with $\vec{r}=\left(x,y,z\left(x,y\right)\right)$. We then assume axial symmetry so that $\vec{r}=(r,\theta,z(r))$. In the small gradient approximation Eq. (2) simplifies to (see Appendix A):
\begin{equation}
\begin{split}
E_{init}\left[z\left(r\right)\right]\approx
&\sum_{i=1,2}\int_{R_{i-1}}^{R_{i}}\pi \kappa_{i}r\big[(\Delta z)^{2} 
-4C_{0i}\Delta z\\
&+\left(2C_{0i}^{2}+ \dfrac{\sigma_i}{\kappa_{i}}\right)(\nabla z)^{2}+\left(4C_{0i}^{2}+2\dfrac{\sigma_i}{\kappa_{i}}\right)\big]dr\\
&+\sum_{i=1,2}\int_{R_{i-1}}^{R_{i}}2\pi\kappa_{G_i}\left(\dfrac{dz}{dr}\dfrac{d^2 z}{dr^{2}}\right)dr\\
&+\sigma_0\left(2\pi\int_0^{R_1}\left(1+\frac{1}{2}(\nabla z_1)^2 \right)rdr-A	 \right)\\
&+\gamma 2\pi R_1,
\end{split}
\end{equation}
where $A$ is the area of domain 1 (the Sla1/Ent1/2 bound component which is attached to the clathrin basket), and $\sigma_0$ is a Lagrange multiplier introduced to fix the area of domain 1. We have also introduced a line tension $\gamma$ at the interface between the two components---the bare component and the Sla1/Ent1/2 bound component~\cite{footnote}. Finally, the radial coordinate of the interface is denoted by $R_1$, while $R_2$ denotes the outer edge of the membrane. We neglect the turgor pressure for now and address it towards the end of this subsection.

We now proceed with the variation of the Lagrangian, $\delta E_{init}\left[z\left(r\right)\right]=\delta\int_0^{R_1}\mathcal{L}_{1}dr
+\delta\int_{R_1}^{R_2}\mathcal{L}_{2}dr=0$ (see Appendix A). As for boundary conditions we demand that membrane be continuous at the interface between the two components so that $z_1(R_1)=z_2(R_1)=z(R_1)$. In addition, the membrane cannot have ridges for the bending energy to remain finite so that $z_1{}'(R_1)=z_2{}'(R_1)=z{}'(R_1)$. For $R_2\rightarrow+\infty$, we have $z_2(R_2)=0$ so that $z_2{}'(R_2)=0$. The radial symmetry demands that $z_1{}'(0)=0$. With these conditions, $\delta z_1, \delta z_2, \delta R_1, \delta z{}'(R_1)$, and $\delta z_1(0)$ are free variables, so the energy is minimized by solving the following equations: 

\beq
\frac{\partial \mathcal{L}_{1}}{\partial z_1}-\frac{d}{dr}\frac{\partial \mathcal{L}_{1}}{\partial z_1{}^\prime}+\frac{d^2}{d r^2}\frac{\partial \mathcal{L}_{1}}{\partial z_1{}^\prime{}^\prime}=0
\eeq
\beq
\frac{\partial \mathcal{L}_{2}}{\partial z_2}-\frac{d}{dr}\frac{\partial \mathcal{L}_{2}}{\partial z_2{}^\prime}+\frac{d^2}{d r^2}\frac{\partial \mathcal{L}_{2}}{\partial z_2{}^\prime{}^\prime}=0
\eeq
\begin{eqnarray}\nonumber
\left.\left[\mathcal{L}_{1}-z_1{}'\left(\frac{\partial \mathcal{L}_{1}}{\partial z_1{}^\prime}-\frac{d}{dr}\frac{\partial \mathcal{L}_{1}}{\partial z_1{}^\prime{}^\prime}\right)-z_1{}''\frac{\partial \mathcal{L}_{1}}{\partial z_1{}^\prime{}^\prime}\right]\right|_{r=R_1}\\
=\left.\left[\mathcal{L}_{2}-z_2{}'\left(\frac{\partial \mathcal{L}_{2}}{\partial z_2{}^\prime}-\frac{d}{dr}\frac{\partial \mathcal{L}_{2}}{\partial z_2{}^\prime{}^\prime}\right)-z_2{}''\frac{\partial \mathcal{L}_{2}}{\partial z_2{}^\prime{}^\prime}\right]\right|_{r=R_1}
\end{eqnarray}
\beq
\left.\left[\left(\frac{\partial \mathcal{L}_{1}}{\partial z_1{}^\prime}-\frac{d}{dr}\frac{\partial \mathcal{L}_{1}}{\partial z_1{}^\prime{}^\prime}\right)-\left(\frac{\partial \mathcal{L}_{2}}{\partial z_2{}^\prime}-\frac{d}{dr}\frac{\partial \mathcal{L}_{2}}{\partial z_2{}^\prime{}^\prime}\right)\right]\right|_{r=R_1}=0
\eeq
\beq
\left.\left[\frac{\partial \mathcal{L}_{1}}{\partial z_1{}^\prime{}^\prime}-\frac{\partial \mathcal{L}_{2}}{\partial z_2{}^\prime{}^\prime}\right]\right|_{r=R_1}=0
\eeq
\beq
\left.\left(\frac{\partial \mathcal{L}_{1}}{\partial z_1{}^\prime}-\frac{d}{dr}\frac{\partial \mathcal{L}_{1}}{\partial z_1{}^\prime{}^\prime}\right)
\right|_{r=0}=0.
\eeq
Solving Eqs. $(6)$ and $(7)$ leads to 
\beq
\begin{split}
z_1(r)&=c_1+c_2\log(r)+c_3 I_0(r\xi_1)+c_4 K_0(r\xi_1)\\
z_2(r)&=c_5+c_6\log(r)+c_7 I_0(r\xi_2)+c_8 K_0(r\xi_2),\\
\end{split}
\eeq
where $\xi_1^2=2C_{01}^{2}+ \dfrac{\sigma_1+\sigma_0}{\kappa_{1}}, \xi_2^2=2C_{02}^{2}+ \dfrac{\sigma_2}{\kappa_{2}}$, and $I_0$ and $K_0$ denote the zeroth order modified Bessel functions of the first and second kind. We can then use Eqs. $(8)$ through $(11)$ and the boundary conditions to determine these 8 coefficients and $\sigma_0$, $R_1$. 

\begin{figure}[t]
\begin{center}
\begin{tabular}{c}
\includegraphics[scale=0.35]{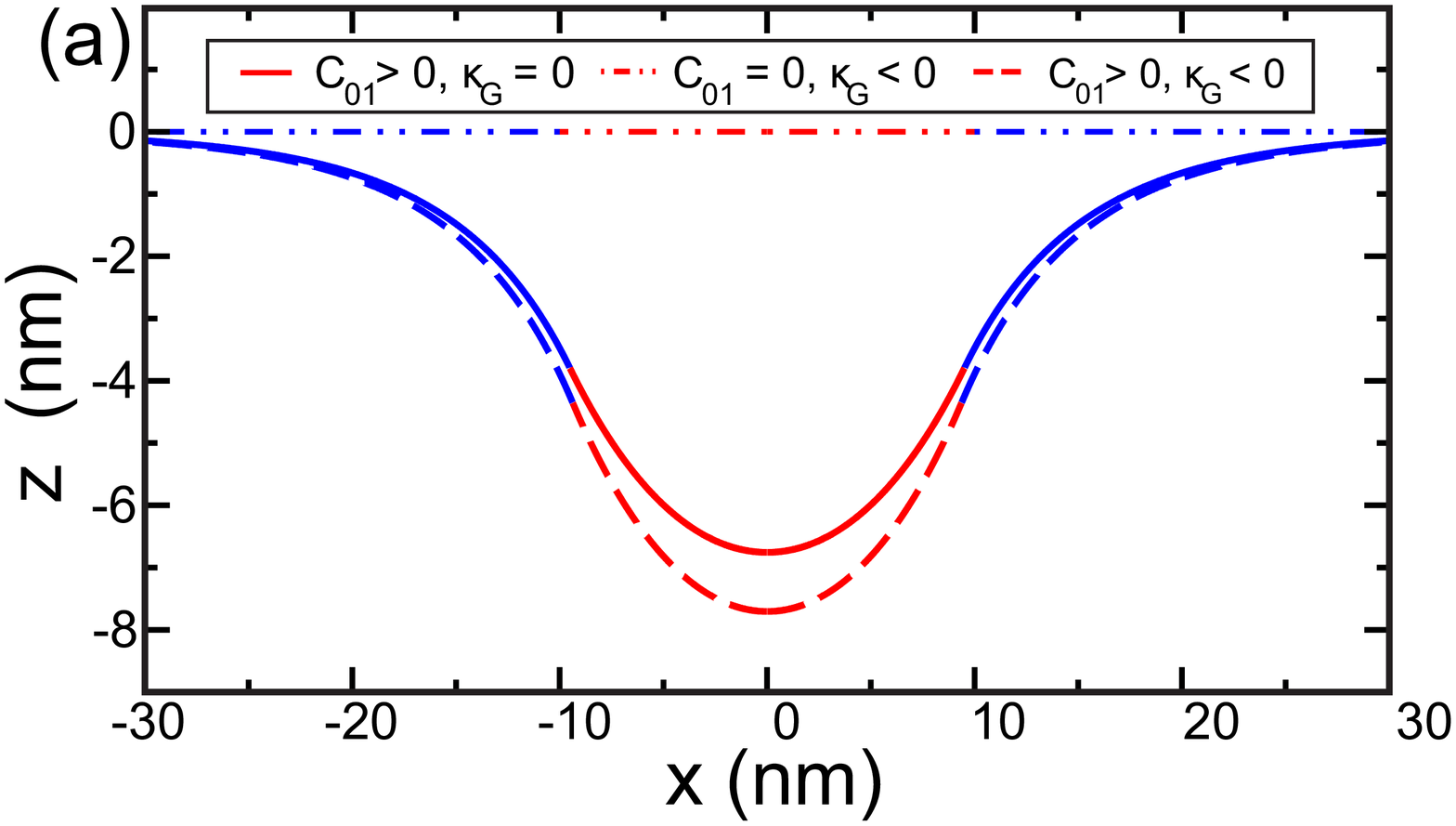}\\
\includegraphics[scale=0.3]{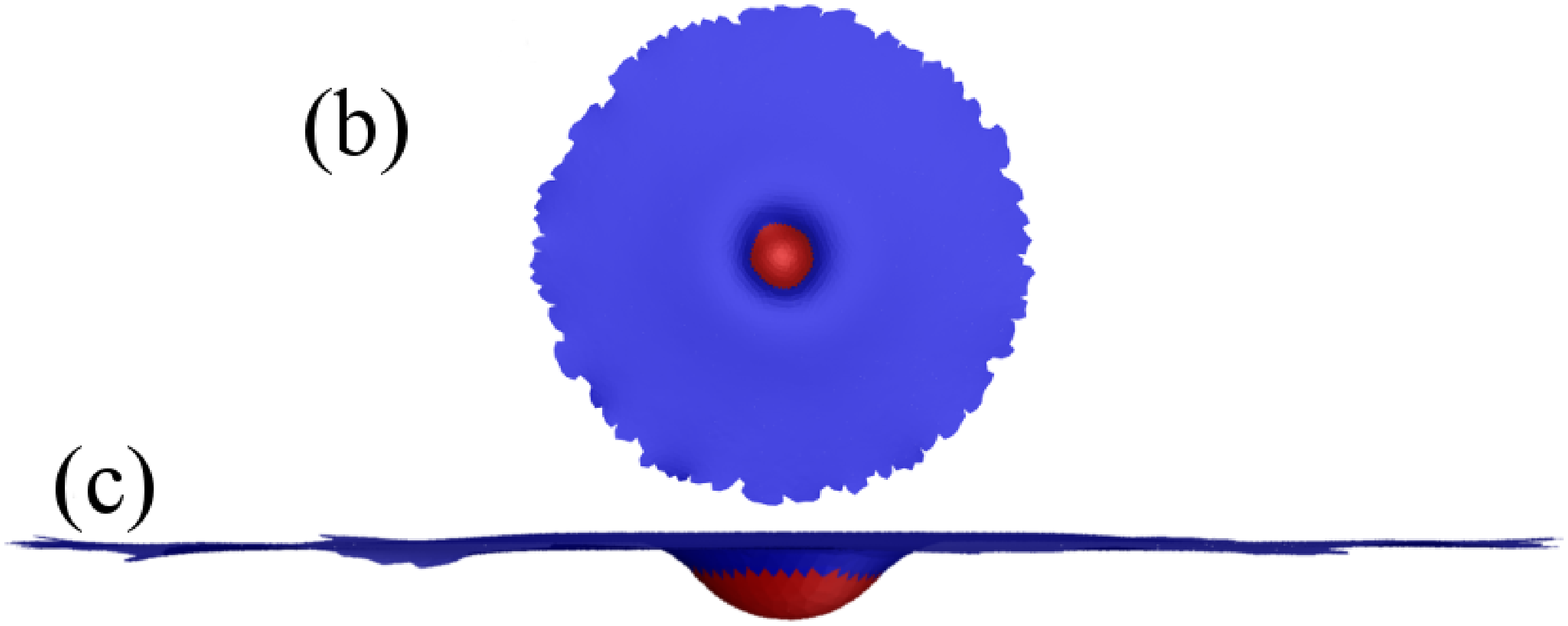}\\
\vspace{0.15cm}\\
\includegraphics[scale=0.35]{initiation_ana_num_compare3.eps}
\end{tabular}
\caption{(a) Cell membrane profile, or $z(x,y=0)$, for the parameters stated in the text. The red(medium grey) denotes the clathrin-bound part of the membrane, while the blue (dark grey) denotes the bare membrane. (b) Top view of the two-component membrane model using simulated annealing Monte Carlo methods. (c) Side view of the same configuration.  Both images have been rescaled accordingly for presentation purposes. (d) Comparison of the maximum depth (or depth) obtained from the numerical simulation (symbols) with the analytical solution (line) for the intiation stage. All the parameters, except for the varying $C_{01}$, are the same as the $\kappa_G=0$ curve in Fig. 4a.}
\end{center}
\end{figure}

For typical parameters for the two-component membrane we use $\kappa_1=20 \,\,k_B T$~\cite{agrawal}, $\kappa_2=10 \,\,k_B T$~\cite{boal}, $C_{01}=0.1 \,\,nm^{-1}$~\cite{ford}, $C_{02}=0$, $\kappa_{G1}=-0.83\kappa_1$~\cite{kozlov}, $\kappa_{G2}=-0.83\kappa_2$, $\sigma_1=0.18\,\, k_B T/ nm^2$~\cite{goud}, $\sigma_2=0.18\,\,k_B T / nm^2$~\cite{goud}, $A=\pi 100\,\,nm^2$, and $\gamma=3\,\,k_B T /nm$~\cite{webb}. Given these parameters, the equilibrium shape of the membrane is plotted in Fig. 4a. We see that a dimple emerges due to clathrin binding.  This dimple finalizes the initiation stage of endocytosis in yeast and sets the radius of the imminent tubular invagination. Fig. 4a contains two additional curves to assess the effect of non-zero Gaussian rigidity $\kappa_G$.  Prior work has found that differences in $\kappa_G$ across a boundary can drive tube formation in membranes~\cite{allain}.  We find, given the above parameters, that it is the non-zero spontaneous curvature that instead drives the dimple formation.  More precisely, for $C_{01}=0$, no dimple forms and for non-zero $C_{01}$ and $\kappa_G$, the depth of the dimple is enhanced only by about 14 percent.

Before concluding this subsection, we must point out that our method differs from an earlier two-component ``dimple'' analysis~\cite{phillips}.  First of all, Ref.~\cite{phillips} does not take into account the $C_{01}^2 (\nabla z)^2$ term~\cite{gozdz}, which is needed for consistency in the small gradient expansion. Second, we take into account non-zero $\kappa_G$ because $\kappa_1\ne\kappa_2$. Thirdly, to solve for some of the coefficients, the earlier work imposes mechanical equilibrium conditions, as opposed to implementing boundary conditions in the variation of the Lagrangian directly. 

Let us now address the presence of turgor pressure.  Previous models use anywhere from $10^3$ Pa~\cite{endo1b,endo1c} to $10^5$ Pa~\cite{bayly} since the turgor pressure at an endocytotic site has not been measured directly. The turgor pressure could be lowered locally by a release of osmolytes near the endocytotic site {it as proposed in Ref.~\cite{bayly} based on experiments presented in Ref.~\cite{tamas}}. For the above set of parameters, using a turgor pressure $p=10^4$ Pa, the energy contribution from the turgor term is 20 percent of the total energy in the absence of the turgor pressure so that the depth of the dimple will decrease slightly. For $p=10^3$ Pa, the turgor energy is only 2 percent of the total energy in the absence of the turgor pressure can be neglected, while for $p=10^5$, a different parameter range would need to be explored, such as the nonzero value of the spontaneous curvature of the membrane whose effect will be described below. Of course, even in the absence of turgor pressure, a depth of 7.7 nm is small such that it may be difficult to measure given an EM pixel size of 2.53 nm~\cite{kukulski}.  The presence of $10^4$ Pa turgor pressure further decreases this depth. Thus, the membrane may not be perfectly flat when the actin filaments begin to polymerize as speculated in ~\cite{kukulski}. 

We also conduct numerical minimization of the intiation stage (Figs. 4b and c). As a check on our simulations, we compare the maximum depth of the dimple, i.e. $|z_{max}(r)|$, as a function of $C_{01}$, for the analytical calculation with the numerical one in Fig. 4d. Here, each energy relaxation is performed starting from a flat configuration.
We place a flat patch of radius $R_2=40l_{0}$ on a hard plane parallel
to the $xy$ plane. We then assign spontaneous curvature $C_{01}$
to all vertices in the central region of radius $r_{C_{01}}=6l_{0}$. We use the energy functional in Eq. (2) with an additional interfacial line tension. Fig. 4d shows the output of the simulation for the same parameters used in Fig. 4a, except with additional $C_{01}$ values. We find very good agreement between the two. Note that since non-zero $\kappa_G$ does not drive the dimple formation (for the parameters studied), we do not include it here.

\subsection{Elongation stage via actin polymerization}
  
Now that the initial small deformation due to Sla1/Ent1/2 binding the membrane to the clathrin basket is formed, the overall radius of the tubular invagination is set. The protein Sla2 next binds to and near the clathrin dimple. Only those Sla2 molecules bound near the top of the clathrin dimple also bind to the actin to form a ring-like structure of binding sites~\cite{lemmon}.  Since the clathrin dimple is elastic-like, it impedes motion of the Sla2 molecules near the top of the clathrin dimple such that these Sla2 binding sites provide for an anchoring of the actin filaments to the membrane to which a localized force can be exerted. As the actin filaments polymerize, the interaction of the polymerizing actin filament tips with the membrane is much more dynamic than the anchoring points since actin filaments polymerize via a ratcheting effect.  The membrane just provides a constraint for the growing filament tips to ratchet against along the length of the tube. This asymmetry in the force is needed for a deformation (other than due to a random fluctuation) to occur. So we model the effect of the anchored at one end, and polymerizing at the other, actin filaments as a localized force on the membrane via the potential $V_{actin}$, as indicated in Eq. (3). In addition, the steric potential $V_{ster}$ models the accumulation of the yeast actin cytoskeleton just beneath the cell membrane and near the tubular invagination as it emerges~\cite{endo3,moseley}. 

How large is this force? An estimate may be obtained from quantitative confocal microscopy measurements of 16 fluorescently labelled proteins involved in endocytosis in fission yeast~\cite{berro1}. The mean peak for the number of G-actin molecules (monomeric actin) is approximately 7500.  Assuming all of these molecules polymerize to form actin filaments of about $100\,\, nm$ in length~\cite{berro2} and each G-actin molecule is $2.7\,\, nm$ in length ($5\,\, nm$ in diameter) then each filament contains about 40 molecules. About 200 actin filaments would then be formed.  Each actin filament contributes approximately $1\,\, pN$ of force,  since the stalling force of an individual actin filament is approximately $1\,\, pN$~\cite{kovar}.  The total force is then approximately $200\,\, pN$, which is applied to the anchoring region of the actin filaments. Since we do not take into account dynamics explicitly, we will merely implement the final value of the total force rather than increasing the force as the actin network develops.  In the quasistatic limit the two approaches should be equivalent. 
  
We now turn to the direction of the actin polymerization force and review the three different proposals for the actin filament orientation~\cite{endo1c,endo4,endo3}. As shown in Fig. 2, actin filaments polymerize ``upward'' and branch via Arp2/3 to drive the membrane further into the interior of the cell.  In Proposal 1, the force is predominantly downward, as opposed to radially inward, provided the initial actin filaments are aligned less than 45 degrees to the normal of the undeformed membrane. Assuming the orienation of the anchored actin filaments stays relatively fixed as branched actin filaments are generated, then the magnitude of the total actin force increases, whilst remaining fixed in direction. Here we assume axial symmetry so that there are only two non-zero components of the total force.  

How do the competing Proposals 2 and 3 compare with Proposal 1? In Proposal 2 the actin network grows inward from the outerlying cytoskeleton towards the invagination site. The actin network is anchored to the outerlying cytoskeleton, as opposed the cell membrane, via Sla2. If we assume a purely radially inward force then the membrane will not deform into a tube with the observed length-to-radius ratio of approximately 10. The branched structure of the actin network, however, can provide a downward component to the total force to elongate the tube (Fig. 3a). We, therefore, distinguish Proposal 1 as a case where the downward component of the total polymerization force is smaller than the radially inward component, whilst in Proposal 2, it is the reverse.

Proposal 3 assumes that there are two anchoring zones for actin filaments---one towards the bottom of the emerging tube and another one near the top of the tube. From these two anchoring zones emerge two actin networks simultaneously growing towards each other and, thus, repelling each other since the actin filaments cannot interpenetrate (Fig. 3b).  It is this repulsion that presumably elongates the tube. Coexistence of a downward force component and an upward force component, however, demands that the membrane simply stretches like a rubber band with no new cell membrane material being added to the tube. Because the cell membrane is bending dominated (and not stretching dominated), Proposal 3 would presumably lead to rupture of the tube~\cite{evans}.   It is not as likely that Proposal 3 contributes to membrane tube formation and we do not study it further as an elongation mechanism.

{\it So we focus on Proposals 1 and 2 for the elongation stage by promoting them to Models 1 and 2, respectively, and study them quantitatively.} To gain some insight, we first review a slightly simpler model, again, first in the absence of turgor pressure. Consider a bare (one-component) membrane with downward force $F$ applied just to the origin, as opposed to being applied over an extended region of the membrane~\cite{prost}.  Assume the membrane has bending rigidity $\kappa$ and surface tension $\sigma$. For a cylindrically shaped membrane with a length $L$ and radius $R$, surface tension favors reducing the radius of the cylinder/tube, while bending favors a larger radius.  Upon minimizing the energy, one obtains an equilibrium radius of $R_{eq}=\sqrt{\kappa/2\sigma}$ and an equilibrium force of $F_{eq}=2\pi\sqrt{2\sigma \kappa}$. For forces less than $F_{eq}$, the membrane deformation is a wide-necked depression and reaches some equilibrium depth that depends on the force, while for forces greater than $F_{eq}$ there is a first-order transition to a cylindrical shape of arbitrary length. Simulations of the membrane shape equation indicate that there is a force barrier from the wide-necked depression to cylinder formation that is 13 percent larger than $F_{eq}$~\cite{prost}. Barriers are a characteristic signature of first-order transitions. Monte Carlo simulations of pure downward pulling on a membrane over an extended region (as opposed to a single point) support this scenario with the force barrier increasing linearly with the size of the region over which the force is exerted~\cite{frenkel}.

Now consider Models 1 and 2 with an additional radially inward force and a steric interaction between the membrane and the actin. To begin, we expect the radially inward force to increase the force barrier to arbitrarily long cylindrical formation. We also expect the steric interaction to alter the transition since $R_{ap}$ cuts off the wide-neck depression and makes it easier to cross-over to long cylinder formation.  More specifically, we expect that as $R_{ap}$ decreases, the change from non-cylindrical to cylindrical occurs at a lower applied force. This effect has been observed in Monte Carlo simulations of driving fluid vesicles through a pore~\cite{pore}. 

To test these notions we study the extension-force curve of tube/cylinder formation for the various models. We do so numerically because $V_{act}$ (Eq. (3)) is a potential localized to a particular region of the membrane, which would be difficult to handle analytically.  We first apply a purely downward force to a ring of vertices right above the Sla1/Ent1/2 attached part of the membrane. We dub this model, Model 0. The magnitude of the total force is denoted by $F_t$ and is distributed uniformly among the vertices.  Since $\kappa_1=10\,\, k_bT$ and $\sigma=0.18 k_B\,\, T/nm$, $F_{eq}\approx 49\,\, pN$(for the applied point force). To numerically determine $F_{eq0}$, the $F_{eq}$ equivalent for Model 0, we pull on the ring with initially $50\,\, pN$ of total force, $F_t$, and $R_{ap}=15\,\, nm$. We then reinitialize $F_t$ to take on smaller values and look for the $F_t$ at the boundary between tubes becoming shorter and tubes becoming longer. We find that $F_{eq0}\approx 25\,\,pN$ using this algorithm.  To study the force barrier, we find that deformations for $F_t<30\,\,pN$ are reasonably robust to perturbations (stepping $F_t$ up and back down again) such that $30 \,\, pN$ is a lower bound for the barrier.  See Fig. 5b.

We now add a radially inward force component to the force applied to the ring of vertices to address Models 1 and 2. How does this radially inward modify the shape crossover due to the downward component of the force? For a purely radially inward force applied to the ring of vertices, the membrane will pucker inward where the force is applied and no cylindrical tube will form. The additional radially inward force increases the force barrier to long tube formation. In Model 1 the actin filaments are anchored at the bottom of the tube so that the downward component of the force is larger than the radially inward component. We assume that $\vec{F}_{t}=\sqrt{2/3}F_{t}\vec{e}_{z}+\sqrt{1/3}F_{t}\vec{e}_{r}$, which would correspond to actin filaments anchoring at an angle of approximately 35 degrees with respect to the normal of the flat part of the membrane. In Model 2 the radially inward component of the actin polymerization force is larger than the downward component, so we choose $\vec{F}_{t}=\sqrt{1/3}F_{t}\vec{e}_{z}+\sqrt{2/3}F_{t}\vec{e}_{r}$. 

Fig. 5c depicts the depth-versus-total force curve for both models as well as two membrane configurations for Model 1. The tubes are reasonably robust to perturbations for all forces studied suggesting that the cross-over to long cylinders is not made. Even though the downward component of the force is increased, it is not enough to overcome the increasing force barrier introduced with the increasing radially inward force as well. For large enough radially inward forces, the tube depth begins to decrease resulting in a nonmonotonic depth-versus-total force curve. As the contribution of the radially inward force increases, it is energetically more favorable for the membrane to deform inward as opposed to elongate. Fig. 5d depicts two equilibrium configurations for Model 1. For the values of $R_{ap}$ studied, $12-18\,\,nm$, the tube depth increases only by several percent with increasing $R_{ap}$. In other words, the applied force clearly plays the dominant role. 

Now on to Model 2. All tubes are reasonably robust to force perturbations, just as in Model 1, and the depth-versus-total force curve is also not monotonic. The largest depth of the membrane deformation for Model 2 is about $65\,\, nm$. While there is indeed some room to play with the ratio of the magnitude of the two components, we contend that Model 1 may better account for the range of observed tube depths~\cite{kukulski}. Hindsight tells us that Model 1 would be more reasonable in obtaining longer tubes, but such depths could have been much longer than the observed ones.  The nonmonotinicity suggests an optimal force of around $100\,\,pN$, should long tubes be the optimizing principle.  And while Model 2 may not necessarily act as the initial driving force to elongate the tube, we address an important role for Model 2, and one aspect of Proposal 3, during the final stage of endocytosis. 

To investigate the role of turgor pressure in Models 1 and 2, we find that as the turgor pressure increases to $10^3$ Pa, our previous results are robust. However, for turgor pressures above this value, the depths of tubes (for a given total force) decrases. See Fig. 5c. So the presence of large turgor pressure biases Model 1 even more so. However, for $p=10^4$ Pa, since the largest depth is approximately 55 nm, to account for larger observed depths, one can invoke the presence of myosin I to allow for extra downward force to increase the depth of the tube~\cite{basu,martin}. Myosin I bind the membrane to the actin filaments.  It has been estimated that there are approximately 300 myosin I molecules at each endoyctotic site, each exerting 2 pN of force~\cite{molloy} (assuming myosin I carry the same force generation potential as myosin II) to arrive at a maximal downward force of 600 pN. Such an additional downward force component would allow for long tubes even in the presence of larger ($10^4-10^5$ Pa) turgor pressures.

\begin{figure}
\begin{center}
\includegraphics[scale=0.3]{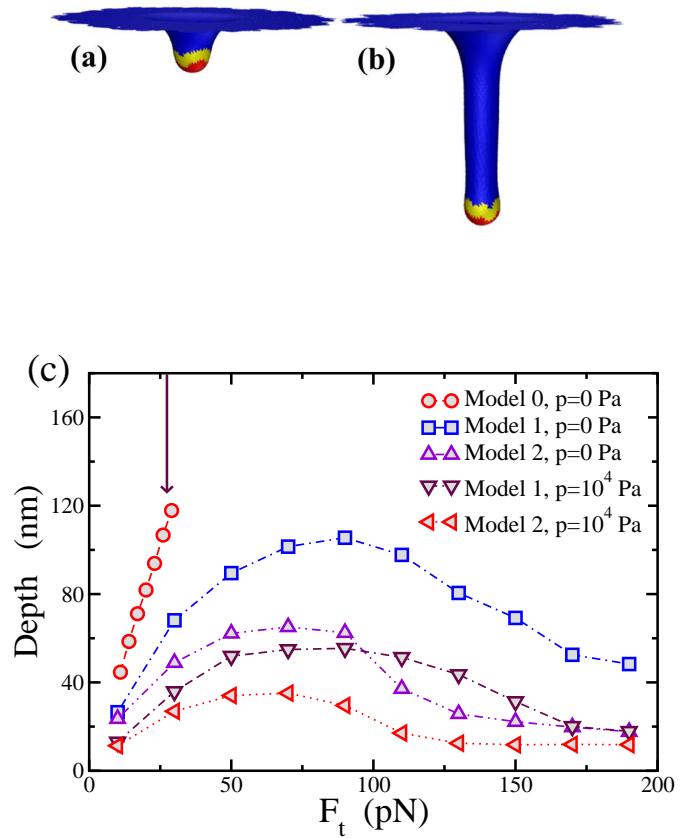}
\includegraphics[scale=0.35]{force.profile.turgor2.eps}
\caption{(a) Simulation results for Model 1 with total applied force $F_{t}=10\,\, pN$. The total force is applied to only the yellow (light grey) part of the membrane (at the vertices). Red (medium grey) denotes the Sla1/Ent1/2 bound part of the membrane and blue (dark grey) denotes the bare membrane. (b) Same as (a) except with $F_{t}=50\,\, pN$; (c) Comparison of the depth as a function of $F_{t}$ for three different models with zero and nonzero turgor pressure, $p$. Again, the error bar is of order the symbol size. The arrow pointing downward denotes the value of $F_{eq}$ for reference .}
\end{center}
\end{figure}

\subsection{Pinch-off stage via the pearling instability}

Experiments indicate that the BAR proteins enter in this last stage, {\it after} the actin filament network has formed~\cite{endo1c,kukulski}. Yet many qualitative depictions of the process show the BAR proteins in between the membrane and the actin filament network~\cite{endo1c}.  BAR proteins have been shown to generate spontaneous curvature in membranes~\cite{sorre}.  Since the BAR proteins enter after the tube has formed~\cite{endo1c,kukulski}, there is no need to generate spontaneous curvature, only sense it. We suggest a potentially new role for BAR proteins here beyond just sensing curvature. Once the tubular-like deformation via the actin filament network occurs, the BAR proteins surround and confine the tube-plus-actin filament network toward the top of part of the tube where bare membrane is exposed to the BAR proteins (Fig. 2d). By surrounding the actin filament network and suppressing the fluctuations of the bare membrane, actin polymerization stops since polymerization is driven by a ratcheting effect in spatially fluctuating fluid membrane (and by the entropically elastic actin network~\cite{mogilneroster}). When actin polymerization stops, no more material can become part of the tube, and the membrane tube area remains constant. 

Because BAR proteins confine part of the actin filament network it is now restricted to lie on the membrane. This effect will generate a new contribution to the membrane energy as indicated in Eq. (4), where the coordinates of the network are the coordinates of the membrane. The actin filament network is modeled as an elastic network with spring constant $\mu$~\cite{brown}. Since actin filaments are semiflexible polymers with a persistence length of about 20 $\mu m$, the elasticity comes from elasticity of the Arp2/3, the branching agent responsible for nucleating new filaments.  The entropic angular spring constant for Arp2/3 is approximately $10^{-19}$ J/rad$^2$~\cite{pollard}, so for branches several actin monomers long, the entropic linear spring constant, $\mu\approx 10^{-2}\,\,N/m$, or $2.5 \,\,k_BT/nm^2$. This additional elasticity contributes to the membrane surface tension with the effective membrane surface tension becoming $\sigma_{eff}=\sigma+\mu/2$, at length scales larger than the meshsize of the actin network. 

How does this increase in surface tension affect the membrane+actin+BAR-protein system? We investigate configurations of a cylindrically shaped membrane with bending rigidity $\kappa$ and increasing surface tension to answer this question. Could such an increase lead to destabilization of the cylindrically-shaped membrane? As the surface tension increases, a sinusoidal perturbation may perhaps lead to the cylindrical membrane breaking up into spherical droplets as surface tension favors spheres. This mechanism is otherwise known as the pearling instability~\cite{pearling,pearling2}.  

Is this instability relevant to the system at hand? Analytical analysis of this instability is included in Appendix B to address this question.  This analysis suggests that the pearling instability may be relevant to the system at hand given the physiological parameters. For the relevant range of wavevectors (less than $0.1\,\,nm^{-1}$), the cylinder is only stable when $\sigma R_o^2/\kappa<2.39$, where $R_o=8.71\,\,nm$, the original radius of the unperturbed cylinder and $\kappa=10\,\,k_BT$. The length-to-radius ratio of the initial cylinder is 10. This inequality, however, depends on the strength of the perturbation.  

To analytically investigate the pearling instability, the volume is assumed to be constant such that the turgor pressure is not important (see Appendix B).  This constraint is imposed so that the membrane does not shrink to a point once the surface tension term dominates.  While, indeed, the invagination is an open system so that the volume of the tube may change slightly, as long as the volume remains finite a pearling instability can set in for some range of parameters. Pearling instabilities have been experimentally observed in open tubes {\it in vivo} and {\it in vitro}~\cite{open1,open2}. 

We implement numerical simulations to numerically test for this instability. We start an initial configuration of a triangulated capsule (as opposed to cylinder) with the above parameters and then vary $\sigma$. As indicated in Fig. 6, the pearling instability mechanism sets in once $\sigma R_0^2/\kappa$ is large enough. In this case, the surface tension must increase by an order of magnitude for the instability to set in.  This increase by an order of magnitude is precisely the contribution of the $E_{act+BAR}$ term in the energy increasing the surface tension from $\sigma\simeq 0.1\,\,k_BT/nm^2$ to $\sigma \simeq 1\,\,k_BT/nm^2$!

\begin{figure}[h]
\centering
\includegraphics[scale=0.25]{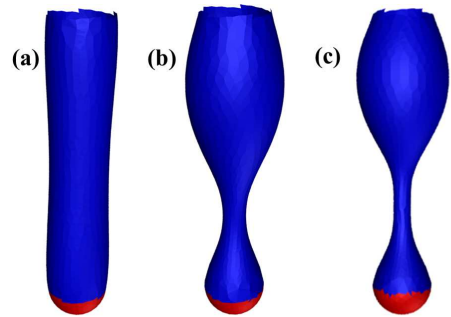}
\caption{The pearling instability for a cylindrical membrane with increasing surface tension going from left to right, or $\sigma R_o^2/\kappa=0.267, 2.67,$ and $4.15$ respectively. The top and red part of the tube are fixed. }
\end{figure}
 
Once the membrane breaks up into spheres, the spheres remain connected, as observed in experiments~\cite{pearling,pearling3}.  This observation differs from the Rayleigh-Plateau instability, where the spheres do not remain connected. So how does the vesicle nearer to the interior of the cell break off from the upper part of the tubular membrane? The most natural answer would be via actin polymerization. Proposals 2 and 3 both provide mechanisms for some downward-directed actin polymerization on the vesicle to drive it further into the cell. In Proposal 2 actin filaments polymerizing inward from the surrounding actin cytoskeleton towards the invagination sites can facilitate break-off of the vesicle by making a comet tail behind it.  Such an actin comet tail has indeed been observed in experiments~\cite{endo3}. In Model 3 anchoring regions of actin filaments to the membrane near the top of the invagination region, while elongation is occuring, could initiate downward actin polymerization. These filaments can then also drive the break off of the vesicle nearer to the interior of the cell from the ``top'' sphere.  Both routes may be important for the final break-off of the vesicle.

\section{Discussion and conclusion}

    We have developed and analyzed a quantitative three-stage model for endocytosis in yeast that is consistent with the experimental data~\cite{endo1}. We first built a model for the initial small membrane deformation due to clathrin indirectly binding to the membrane via Sla1/Ent1/2.  We demonstrated that the Sla1/Ent1/2-bound domain initiates invagination of the membrane by forming a small depression, or dimple, to set the radius of the subsequent tubular invagination. This subsequent tubular invagination is driven by actin polymerization forces, which we model as an external force applied to the membrane. We found that of the three competing proposals in the literature for the orientation of the actin filaments in driving tube formation, one proposal (Proposal 1)~\cite{endo1c} is most likely to account for the observed tubular lengthscales of the cell membrane in endocytosis in yeast~\cite{kukulski}.  For turgor pressures smaller than $10^{3}$ Pa, our results predict the applied force that optimizes the length of the tube, where the largest length-to-radius ratio is approximately 10. For turgor pressures larger than $10^3$ Pa, myosin I, an actin motor that binds directly to the cell membrane so that it can enhance actin-dependent forces on the membrane, can potentially account for the large length-to-radius ratio~\cite{basu}. The combination of this large ratio and the effective surface tension increases due to the presence of BAR proteins confining the actin filament network against the tubular cell membrane (Fig. 2d) naturally motivates that the pearling instability may assist in the scission mechanism. We showed that the pearling instability may promote spherical vesicle formation by both analytical calculations and simulations given the physiological parameters involved in endocytosis in yeast.

Let us contrast our model with an earlier model for endocytosis in yeast~\cite{endo1b,endo1c}. In the latter, the coordinated effect of protein-induced lipid phase segregation along the tubule plays a key role in vesicle scission. The phase separation between hydrolyzed and non-hydrolyzed PIP$_2$, a membrane-bound protein to which actin attaches, calls for a two-component fluid membrane and induces an interfacial line tension between the two components to drive pinch-off. The effect of actin in this model is to decrease the effective surface tension of the membrane, which makes it easier for the interfacial line tension to scission the membrane and is rather different than the effect of actin in our model. We, however, model the actin as an applied force and are able to generate tube formations as a result.  The competing quantitative model is not able to generate tubes explicitly given the manner in which actin polymerization is incorporated into the model. We also demonstrate that the pearling instability could potentially facilitate pinch-off.  The frequency of endocytosis in budding yeast, invaginating its total cell membrane surface in about 100 minutes~\cite{kukulski}, suggests that an instability, as opposed to coordinated effort involving lipid phase separation, would be useful. The observation that scission occurs at a range of invagination depths also favors an instability as opposed to a more regulated mechanism. Comparison with another recent model is rather difficult since this new model assumes that the cell membrane is an elastic membrane with a nonzero shear modulus rather than assuming the cell membrane is a fluid one~\cite{bayly}. The formation of tubes in elastic membranes is very different from the formation of tubes in fluid membranes, where there is no shear modulus.

While we have presented a quantitative model for endocytosis in yeast, how much of this story applies to endocytosis in mammalian cells? More spherical-like membrane deformations are generated in mammalian cells due to clathrin cage formation and the motor protein, dynamin, driving pinch-off? Many of the same proteins involved in yeast clathrin-mediated endocytosis (CME) are conserved in mammalian CME~\cite{perrais}. It could be that the presence of the turgor pressure in yeast makes clathrin cage assembly difficult, but clathrin basket assembly less difficult given the much smaller change in volume for the basket. Then it is up to the actin, etc. to finish the job.  As some of our results depend on the strength of the turgor pressure, it would be good to measure it directly at an endocytotic site. There is also another route to endocytosis in mammalian cells via the CLIC/GEEC pathway, which does not require clathrin or dynamin, and forms more tubular deformations as observed in yeast~\cite{geec}. The requirement for actin in mammalian CME has been less clear. Several new studies in mammalian cells provide support for an actin requirement in the invagination and late stages of CME~\cite{perrais}. On the other hand, a recent {\it in vitro} experiment with clathrin and dynamin suggest that these two proteins are sufficient to drive endocytosis in mammalian cells~\cite{invitro}.  In light of this experiment, it would be interesting to revisit the modeling of endocytosis in mammalian cells~\cite{bruinsma}. It may also be useful to investigate how the modeling presented here can be extended to enveloped virus entry~\cite{sun}, extocytosis, and budding to form a more unified theoretical framework for cell membrane deformations used to transport material in and out of the cell. 

\section{Acknowledgements}
The authors acknowledge support from the Soft Matter Program at Syracuse University and the Aspen Center for Physics, where part of this work was completed. MJB acknowledges support from NSF grant DMR-0808812. The authors also acknowledge helpful discussion with R. Bruinsma and J. Guven.

\onecolumngrid
\appendix

\section{Details of variational calculation for initiation stage in the Monge representation}
To examine the equilibrium shape of the cell membrane with clathrin indirectly attached in a localized region (the invagination site), we consider the following energy~\cite{helfrich}:
\beq
E_{init}=
\sum_{i=1,2}\int dS_i\left[2\kappa_i \left(H_i-C_{0i} \right)^2+\kappa_{G_i} K_i+\sigma_i\right]
\eeq
where $i=1$ denotes the Sla1/Ent1/2-bound membrane (to which the clathrin then attaches) and $i=2$ denotes the bare membrane.  In addition, $\kappa_1$ denotes the bending rigidity of component 1, while $\kappa_2$ denotes the bending rigidity of component 2. The respective spontaneous curvatures are denoted by $C_{01}$ and $C_{02}$ respectively. $\kappa_{G_1}$ denotes the saddle-splay modulus of component 1, while $\kappa_{G_2}$ denotes the saddle-splay modulus of component 2. Finally, $\sigma$ denotes the surface tension. 

We represent the membrane shape using the so-called Monge representation such
 that each point on the membrane in three-dimensional space is given as 
\beq
\vec{r}=\left(x,y,z\left(x,y\right)\right).
\eeq
It is straightforward to show that the differential area element is
\beq
dS=\sqrt{1+\left(\nabla z\right)^{2}}dxdy,
\eeq
and the mean curvature is
\beq
H=\frac{1}{2}\nabla\left(\frac{\nabla z}{\sqrt{1+\left(\nabla z\right)^{2}}}\right).
\eeq

In the small gradient approximation, or $\left|\nabla z\right|\ll1$,  
\beq
dS\approx \left(1+\frac{1}{2}\left(\nabla z\right)^2\right)dxdy,
\eeq
and
\beq
H\approx\frac{1}{2}\Delta z,
\eeq
where $\Delta z=\frac{\partial^{2}z}{\partial x^{2}}+\frac{\partial^{2}z}{\partial y^{2}}$ is the Laplacian. Assuming the system is axisymmetric $z=z\left(\sqrt{x^{2}+y^{2}}\right)\equiv z\left(r\right)$, we write the Laplacian in the cylindrical coordinates as
\beq
\Delta z=\frac{1}{r}\dfrac{dz}{dr}+\dfrac{d^2 z}{dr^{2}},
\eeq
where $r$ is the distance from the origin, which we set to coincide with the center of a circular membrane.

Additionally, the Gaussian curvature is given by 
\beq
K=\frac{1}{r}\dfrac{dz}{dr}\dfrac{d^2 z}{dr^{2}}.
\eeq

The energy can now be written in the approximate form
\beq
\begin{split}
E_{init}\left[z\left(r\right)\right] \approx &\sum_{i=1,2}\int_{R_{i-1}}^{R_{i}}2\pi r\left[2\kappa_{i}\left(\frac{1}{2} \Delta z-C_{0i}\right)^{2}+\kappa_{G_i}\left(\frac{1}{r}\dfrac{dz}{dr}\dfrac{d^2 z}{dr^{2}}\right)+\sigma_{i}\right]\left(1+\frac{1}{2}\left(\nabla z\right)^{2}\right)dr\\
 \approx &\sum_{i=1,2}\int_{R_{i-1}}^{R_{i}}\pi \kappa_{i}r\left[(\Delta z)^{2}-4C_{0i}\Delta z+\left(2C_{0i}^{2}+ \dfrac{\sigma_i}{\kappa_{i}}\right)(\nabla z)^{2}+\left(4C_{0i}^{2}+2\dfrac{\sigma_i}{\kappa_{i}}\right)\right]dr\\
&+\sum_{i=1,2}\int_{R_{i-1}}^{R_{i}}2\pi\kappa_{G_i}\left(\dfrac{dz}{dr}\dfrac{d^2 z}{dr^{2}}\right)dr
\end{split}
\eeq
where $R_0=0$. Next, we fix the area of component 1 (the Sla1/Ent1/2 attached domain) to be area $A$ by introducing a Lagrange multiplier and also consider line tension at the interface between the two components/domains, then 
\beq
\begin{split}
E_{init}\left[z\left(r\right)\right]=&\sum_{i=1,2}\int_{R_{i-1}}^{R_{i}}\pi \kappa_{i}r\left[(\Delta z)^{2}-4C_{0i}\Delta z+\left(2C_{0i}^{2}+ \dfrac{\sigma_i}{\kappa_{i}}\right)(\nabla z)^{2}+\left(4C_{0i}^{2}+2\dfrac{\sigma_i}{\kappa_{i}}\right)\right]dr\\
&+\sum_{i=1,2}\int_{R_{i-1}}^{R_{i}}2\pi\kappa_{G_i}\left(\dfrac{dz}{dr}\dfrac{d^2 z}{dr^{2}}\right)dr+\sigma_0\left(2\pi\int_0^{R_1}\left(1+\frac{1}{2}(\nabla z_1)^2 \right)rdr-A	 \right)
+\gamma 2\pi R_1
\end{split}
\eeq
We define $z'\equiv\frac{dz}{dr}, z''\equiv\frac{d^2z}{dr^2}, \xi_1^2=2C_{01}^{2}+ \dfrac{\sigma_1+\sigma_0}{\kappa_{1}}, \xi_2^2=2C_{02}^{2}+ \dfrac{\sigma_2}{\kappa_{2}}$ such that the Lagrangian for each component becomes 
\beq
\begin{split}
\mathcal{L}_{1}&=\pi\kappa_{1}r\left[\left(\frac{1}{r}z_1{}'+z_1{}''\right)^{2}-4C_{01}\left(\frac{1}{r}z_1{}'+z_1{}''\right)+\xi_1^2(z_1{}')^{2}+2\xi_1^2\right]+2\pi\kappa_{G_1}z_1{}'z_1{}''+2\pi\gamma r \delta(r-R_1)\\
\mathcal{L}_{2}&=\pi\kappa_{2}r\left[\left(\frac{1}{r}z_2{}'+z_2{}''\right)^{2}-4C_{02}\left(\frac{1}{r}z_2{}'+z_2{}''\right)+\xi_2^2(z_2{}')^{2}+2\xi_2^2\right]+2\pi\kappa_{G_2}z_2{}'z_2{}''\\
\end{split}
\eeq

We now proceed with the variation of the Lagrangian, or $\delta E_{init}\left[z\left(r\right)\right]=\delta\int_0^{R_1}\mathcal{L}_{1}dr
+\delta\int_{R_1}^{R_2}\mathcal{L}_{2}dr$. Specifically, we have
\beq
\begin{split}
\delta\int_0^{R_1}\mathcal{L}_{1}dr=&
\left.\left[\mathcal{L}_{1}-z_1{}'\left(\frac{\partial \mathcal{L}_{1}}{\partial z_1{}^\prime}-\frac{d}{dr}\frac{\partial \mathcal{L}_{1}}{\partial z_1{}^\prime{}^\prime}\right)-z_1{}''\frac{\partial \mathcal{L}_{1}}{\partial z_1{}^\prime{}^\prime}\right]\right|_{r=R_1}\delta R_1\\
&+\int_0^{R1}\left(\frac{\partial \mathcal{L}_{1}}{\partial z_1}-\frac{d}{dr}\frac{\partial \mathcal{L}_{1}}{\partial z_1{}^\prime}+\frac{d^2}{d r^2}\frac{\partial \mathcal{L}_{1}}{\partial z_1{}^\prime{}^\prime}\right)\delta z_1dr
+\left.\left(\frac{\partial \mathcal{L}_{1}}{\partial z_1{}^\prime}-\frac{d}{dr}\frac{\partial \mathcal{L}_{1}}{\partial z_1{}^\prime{}^\prime}\right)
\right|_{r=R_1}\delta z_1(R_1)\\
&-\left.\left(\frac{\partial \mathcal{L}_{1}}{\partial z_1{}^\prime}-\frac{d}{dr}\frac{\partial \mathcal{L}_{1}}{\partial z_1{}^\prime{}^\prime}\right)
\right|_{r=0}\delta z_1(0)+\left.\frac{\partial \mathcal{L}_{1}}{\partial z_1{}^\prime{}^\prime}\right|_{r=R_1}\delta z_1{}'(R_1),\\
\end{split}
\eeq
and 
\beq
\begin{split}
\delta\int_{R_1}^{R_2}\mathcal{L}_{2}dr=&
\left.\left[\mathcal{L}_{2}-z_2{}'\left(\frac{\partial \mathcal{L}_{2}}{\partial z_2{}^\prime}-\frac{d}{dr}\frac{\partial \mathcal{L}_{2}}{\partial z_2{}^\prime{}^\prime}\right)-z_2{}''\frac{\partial \mathcal{L}_{2}}{\partial z_2{}^\prime{}^\prime}\right]\right|_{r=R_2}\delta R_2\\
& -\left.\left[\mathcal{L}_{2}-z_2{}'\left(\frac{\partial \mathcal{L}_{2}}{\partial z_2{}^\prime}-\frac{d}{dr}\frac{\partial \mathcal{L}_{2}}{\partial z_2{}^\prime{}^\prime}\right)-z_2{}''\frac{\partial \mathcal{L}_{2}}{\partial z_2{}^\prime{}^\prime}\right]\right|_{r=R_1}\delta R_1\\
&+\int_{R_1}^{R2}\left(\frac{\partial \mathcal{L}_{2}}{\partial z_2}-\frac{d}{dr}\frac{\partial \mathcal{L}_{2}}{\partial z_2{}^\prime}+\frac{d^2}{d r^2}\frac{\partial \mathcal{L}_{2}}{\partial z_2{}^\prime{}^\prime}\right)\delta z_2dr
+\left.\left(\frac{\partial \mathcal{L}_{2}}{\partial z_2{}^\prime}-\frac{d}{dr}\frac{\partial \mathcal{L}_{2}}{\partial z_2{}^\prime{}^\prime}\right)
\right|_{r=R_2}\delta z_2(R_2)\\
&-\left.\left(\frac{\partial \mathcal{L}_{2}}{\partial z_2{}^\prime}-\frac{d}{dr}\frac{\partial \mathcal{L}_{2}}{\partial z_1{}^\prime{}^\prime}\right)
\right|_{r=R_1}\delta z_2(R_1)+\left.\frac{\partial \mathcal{L}_{2}}{\partial z_2{}^\prime{}^\prime}\right|_{r=R_2}\delta z_2{}'(R_2)-\left.\frac{\partial \mathcal{L}_{2}}{\partial z_2{}^\prime{}^\prime}\right|_{r=R_1}\delta z_1{}'(R_1).\\
\end{split}
\eeq

So, 
\beq
\begin{split}
\delta E_{init}\left[z\left(r\right)\right]=
&\int_0^{R1}\left(\frac{\partial \mathcal{L}_{1}}{\partial z_1}-\frac{d}{dr}\frac{\partial \mathcal{L}_{1}}{\partial z_1{}^\prime}+\frac{d^2}{d r^2}\frac{\partial \mathcal{L}_{1}}{\partial z_1{}^\prime{}^\prime}\right)\delta z_1dr
+\int_{R_1}^{R2}\left(\frac{\partial \mathcal{L}_{2}}{\partial z_2}-\frac{d}{dr}\frac{\partial \mathcal{L}_{2}}{\partial z_2{}^\prime}+\frac{d^2}{d r^2}\frac{\partial \mathcal{L}_{2}}{\partial z_2{}^\prime{}^\prime}\right)\delta z_2dr\\
&+\left.\left[\mathcal{L}_{1}-z_1{}'\left(\frac{\partial \mathcal{L}_{1}}{\partial z_1{}^\prime}-\frac{d}{dr}\frac{\partial \mathcal{L}_{1}}{\partial z_1{}^\prime{}^\prime}\right)-z_1{}''\frac{\partial \mathcal{L}_{1}}{\partial z_1{}^\prime{}^\prime}\right]\right|_{r=R_1}\delta R_1\\
&-\left.\left[\mathcal{L}_{2}-z_2{}'\left(\frac{\partial \mathcal{L}_{2}}{\partial z_2{}^\prime}-\frac{d}{dr}\frac{\partial \mathcal{L}_{2}}{\partial z_2{}^\prime{}^\prime}\right)-z_2{}''\frac{\partial \mathcal{L}_{2}}{\partial z_2{}^\prime{}^\prime}\right]\right|_{r=R_1}\delta R_1\\
&+\left.\left[\mathcal{L}_{2}-z_2{}'\left(\frac{\partial \mathcal{L}_{2}}{\partial z_2{}^\prime}-\frac{d}{dr}\frac{\partial \mathcal{L}_{2}}{\partial z_2{}^\prime{}^\prime}\right)-z_2{}''\frac{\partial \mathcal{L}_{2}}{\partial z_2{}^\prime{}^\prime}\right]\right|_{r=R_2}\delta R_2\\
&+\left.\left[\left(\frac{\partial \mathcal{L}_{1}}{\partial z_1{}^\prime}-\frac{d}{dr}\frac{\partial \mathcal{L}_{1}}{\partial z_1{}^\prime{}^\prime}\right)
-\left(\frac{\partial \mathcal{L}_{2}}{\partial z_2{}^\prime}-\frac{d}{dr}\frac{\partial \mathcal{L}_{2}}{\partial z_2{}^\prime{}^\prime}\right)\right]
\right|_{r=R_1}\delta z(R_1)\\
&+\left.\left[\frac{\partial \mathcal{L}_{1}}{\partial z_1{}^\prime{}^\prime}-\frac{\partial \mathcal{L}_{2}}{\partial z_2{}^\prime{}^\prime}\right]\right|_{r=R_1}\delta z{}'(R_1)
-\left.\left(\frac{\partial \mathcal{L}_{1}}{\partial z_1{}^\prime}-\frac{d}{dr}\frac{\partial \mathcal{L}_{1}}{\partial z_1{}^\prime{}^\prime}\right)
\right|_{r=0}\delta z_1(0)\\
&+\left.\frac{\partial \mathcal{L}_{2}}{\partial z_2{}^\prime{}^\prime}\right|_{r=R_2}\delta z_2{}'(R_2)+\left.\left(\frac{\partial \mathcal{L}_{2}}{\partial z_2{}^\prime}-\frac{d}{dr}\frac{\partial \mathcal{L}_{2}}{\partial z_2{}^\prime{}^\prime}\right)
\right|_{r=R_2}\delta z_2(R_2),\\
\end{split}
\eeq
where we assume $z_1(R_1)=z_2(R_1)=z(R_1)$ and $z_1{}'(R_1)=z_2{}'(R_1)=z{}'(R_1)$.

\section{Pearling instability analysis}
To understand how the pearling instability comes about as surface tension increases, we model the tubular invagination as a cylinder with a surface of revolution along the $z$ axis given in a parametric form 
\begin{equation}
\vec{r}=\left(f\left(z\right)\cos\varphi,f\left(z\right)\sin\varphi,z\right),\label{eq:revolution}
\end{equation}
where $f\left(z\right)$ has the form 
\beq
f\left(z\right)=R\left(1+\frac{a}{R}\zeta\left(z\right)\right),
\eeq
where $\frac{a}{R}\ll1$ and $\left|\zeta\left(z\right)\right|\sim1$ with the assumption $\zeta\left(0\right)=\zeta\left(L\right)=0$. See Fig. 7a.
 The simplest possible form for $\zeta$ is 
\beq
\zeta\left(z\right)=\sin\left(qz\right).
\eeq
Moreover, to quantify the energy, the mean curvature of the surface of revolution~\cite{revolution} is
\beq
\begin{split}
H\left(z\right) & = \frac{f(z)f''(z)-f'(z)^{2}-1}{2f(z)\left(f'(z)^{2}+1\right)^{3/2}}\\
& = \frac{-1-a^2 q^2 \cos(q z)^2-a q^2 \sin(q z) (R+a \sin(q z))}{2 \left(1+a^2 q^2 \cos(q z)^2\right)^{3/2} (R+a \sin(q z))}\\
\end{split}
\eeq

In this parameterization, the energy of the cylindrical membrane is given as
\beq
\begin{split}
E & =\int dS \left( 2\kappa H^{2}+\sigma\right)\\
&=\int d\varphi dzf\left(z\right)\left(1+f'\left(z\right)^{2}\right)^{1/2}\left( 2\kappa H^{2}+\sigma\right).
\end{split}
\eeq
This expands to
\begin{eqnarray}
E=\int d\varphi dzR\left(1+\frac{a}{R}\zeta\left(z\right)\right)\left(1+a^{2}\zeta'\left(z\right)^{2}\right)^{1/2}\left( 2\kappa H^{2}+\sigma\right)\nonumber\\
= \int_0^L 2 \pi  \sqrt{1+a^2 q^2 \cos(q z)^2} (R+a \sin(q z))\big(\sigma \nonumber\\
+\frac{\kappa  \left(1+a^2 q^2 \cos(q z)^2+a q^2 \sin(q z) (R+a \sin(q z))\right)^2}{2 \left(1+a^2 q^2 \cos(q z)^2\right)^3 (R+a \sin(q z))^2}\big)dz.
\end{eqnarray}
We will now impose a volume constraint as with all minimal surface problems, otherwise, once the surface tension term dominates, no surface will be the lowest energy solution. This volume constaint is given by 

\beq
V=\int_0^L \pi\left(R+a\sin\left(qz\right) \right)^2 dz=\pi R_o^2 L, 
\eeq
where $R_o$ is the original cylinder's radius. Using this constraint, we can solve for $R$ and then compare the energies of the original unperturbed cylinder with the perturbed cylinder, where the former energy, $E_0$, is given by  
\beq
\begin{split}
E_0 & =E(a=0)=\int_0^L 2 \pi  R_o \left(\dfrac{\kappa }{2 R_o^2}+\sigma \right)dz \\
 & =2 \pi R_o L \left(\dfrac{\kappa }{2 R_o^2}+\sigma \right).
\end{split}
\eeq
Using {\it Mathematica}~\cite{mathematica}, we find that when $\kappa=10\,\,k_B T$, $\sigma=0.32\,\,k_B T/nm^2$, $R_o=8.71 \,\,nm$, $L=87.1\,\,nm$---all relevant parameters for the problem at hand---and $a=0.16\,\,nm$, the energy difference $\Delta E=E_{pert}-E_0$ becomes negative near $q\approx 0.06\,\,nm^{-1}$ (See Fig. 7b). For a cylinder of order $100 \,\,nm$, this instability can lead to pinch-off. We must point out that for $\sigma R_o^2/\kappa>1.6$, there is an instability at much smaller $q$, or much longer cylinders~\cite{witten}, such that this instability would not be relevant for endocytosis in yeast. For the relevant range of wavevectors, for $\sigma R_o^2/\kappa<2.39$, the smooth cylinder is stable.  Therefore, the ratio $\sigma R_o^2/\kappa$ needs to be large enough for the pearling instability to set in for endocytosis in yeast since surface tension favors spheres as opposed to cylinders. In other words, once the surface tension becomes large enough, the pearling instability sets in. 

\begin{figure}[h]
\begin{center}
\includegraphics[scale=0.5]{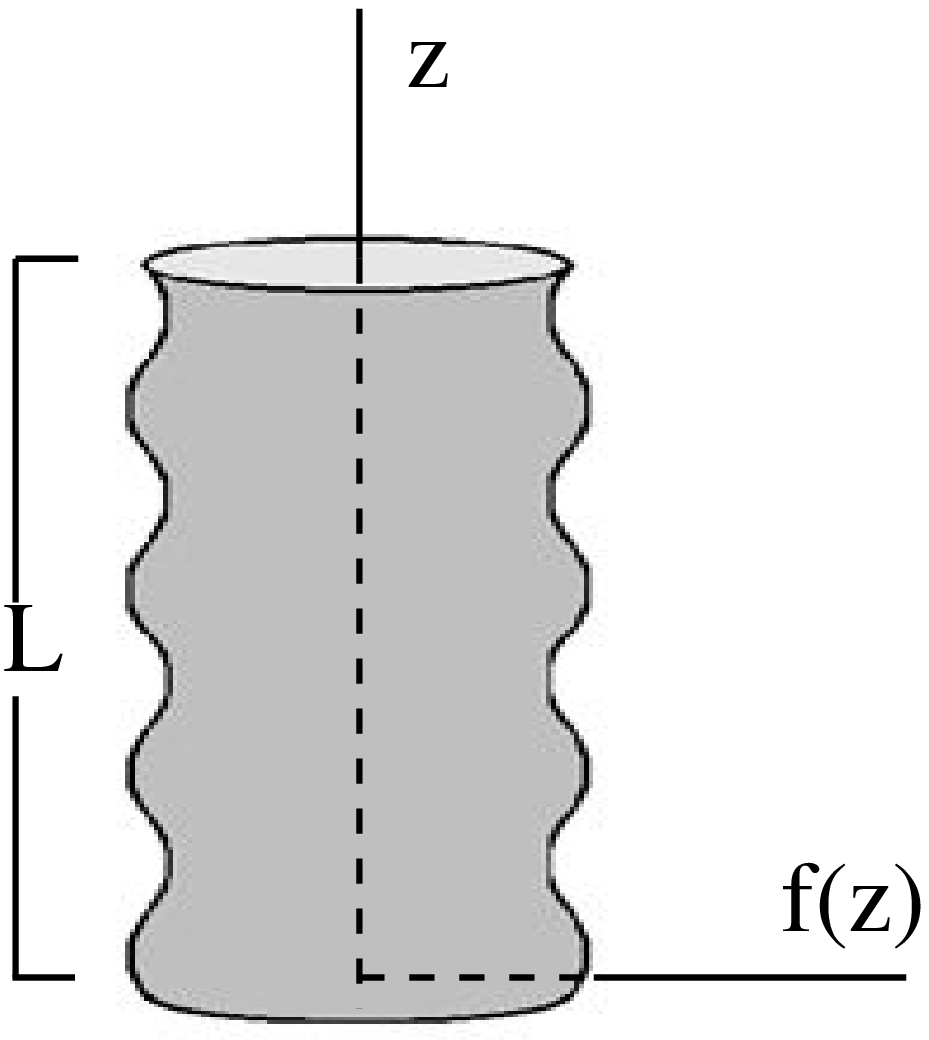}
\hspace{1cm}
\includegraphics[scale=0.35]{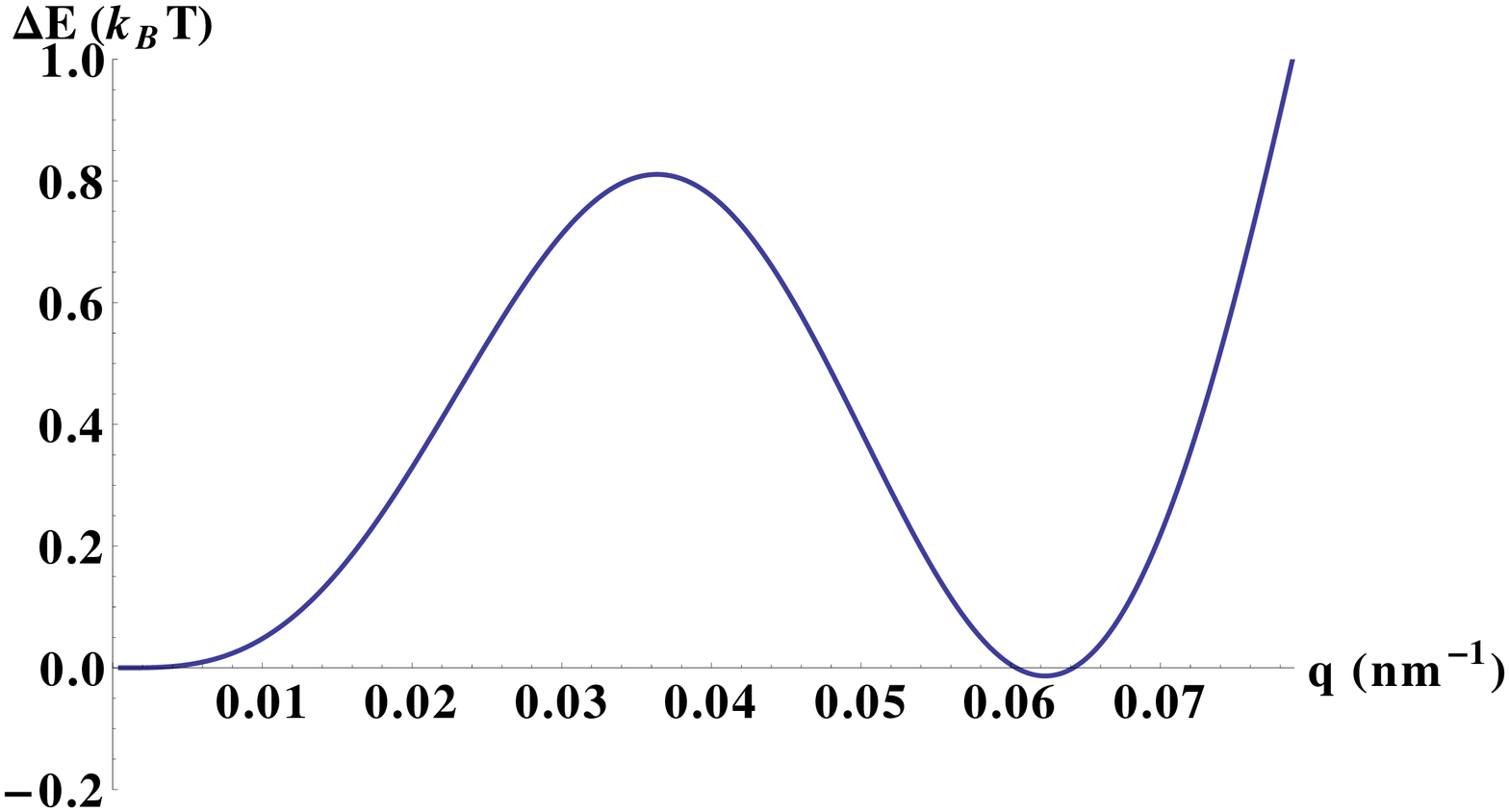}
\caption{Left: Schematic denoting notation used. Right: Difference in energy between the perturbed and unperturbed cylinder as a function of wavenumber $q$ for $\sigma R_o^2/\kappa=2.67$, $L/R=10$, and $a=0.16\,\,nm$.}
\end{center}
\end{figure}


\begin{thebibliography}{99}
\bibitem{footnote} Note that Ent1/2 are the yeast homologues for epsin in mammalian cells.
\end{thebibliography}
\end{document}